\documentclass[structabstract]{aa}  
\usepackage{graphicx,natbib}
\usepackage{txfonts}
\usepackage{longtable}

\newcommand{\kms}{~km~s$^{-1}$}
\newcommand{\teff}{$T_{\rm eff}$}
\newcommand{\logg}{$\log g$}

\newcommand{\ebv}{$E(B-V)$}
\begin{document}
   \title{Chemical compositions of six metal-poor stars in the
   ultra-faint dwarf spheroidal galaxy Bo\"{o}tes I}


   \author{M. N. Ishigaki\inst{1} 
         \and
          W. Aoki\inst{2,3}
          \and
          N. Arimoto\inst{2,3,4}
          \and
          S. Okamoto\inst{5}
          }

   \institute{
Kavli Institute for the Physics and Mathematics of the Universe (WPI), Todai Institute for Advanced Study, University of Tokyo, 5-1-5 Kashiwanoha, Kashiwa, Chiba 277-8583, Japan \email{miho.ishigaki@ipmu.jp}
     \and
National Astronomical Observatory of Japan, Mitaka, Tokyo 181-8588, Japan 
              \email{aoki.wako@nao.ac.jp, arimoto.n@nao.ac.jp}
         \and
Department of Astronomical Science, The Graduate University of Advanced Studies, Mitaka, Tokyo 181-8588, Japan
\and
Subaru Telescope, National Astronomical Observatory of Japan, 650 North A’ohoku Place, Hilo, HI 96720, USA
         \and
             Kavli Institute for Astronomy and Astrophysics, Peking University, Beijing 100871, China 
             \email{okamoto@pku.edu.cn}
         }

   \date{}

 
  \abstract
  { Ultra-faint dwarf galaxies recently discovered around
  the Milky Way (MW) contain extremely metal-poor stars, and might
  represent the building blocks of low-metallicity components of
  the MW.  Among them, the Bo\"{o}tes I dwarf spheroidal galaxy 
is of particular interest because of its exclusively old stellar population. 
Detailed chemical compositions of individual stars in
    this galaxy are a key to understanding formation and 
chemical evolution in the oldest galaxies in the Universe and 
their roles in building up the MW halo. 
}
  {Previous studies of the chemical abundances 
of Bo\"{o}tes I show discrepancies in elemental abundances between 
different authors, and thus a 
consistent picture of its chemical enrichment history 
has not yet been established. 
In the present work, we independently determine chemical compositions of 
 six red giant stars in Bo\"{o}tes I, some of which overlap with those 
analyzed in the previous 
studies. Based on the derived abundances, we re-examine 
trends and scatters in 
elemental abundances and make comparisons with MW field halo stars 
and other dwarf spheroidal galaxies in the MW.    
 }
  {High-resolution spectra of a sample of stars were 
obtained with the High Dispersion 
Spectrograph mounted on the Subaru
    Telescope. Abundances of 12 elements, including C, Na, $\alpha$, 
Fe-peak, and neutron capture elements, were determined for the sample stars. 
The abundance results were compared to those 
    in field MW halo stars previously obtained 
  using an abundance analysis technique
    similar to the present study.}
   { We confirm the low metallicity of Boo-094 ([Fe/H]$=-3.4$). 
Except for this star, the abundance ratios ([X/Fe]) 
of elements lighter than zinc 
are generally homogeneous with small scatter around the mean values 
in the metallicities 
spanned by the other five stars ($-2.7<$[Fe/H]$<-1.8$). 
Specifically, all of the 
sample stars with [Fe/H]$>-2.7$ 
show no significant enhancement of carbon. 
The [Mg/Fe] and [Ca/Fe] ratios are almost constant with 
a modest decreasing trend with 
increasing [Fe/H] and are slightly lower than the field halo stars. 
The [Sr/Fe] and [Sr/Ba] ratios 
also tend to be lower in the Bo\"{o}tes I stars than in the halo stars.} 
{Our results of small scatter in the [X/Fe] ratios for elements lighter than 
zinc suggest that these abundances were homogeneous among the 
ejecta of prior generation(s) of stars in this galaxy.   
The lower mean [Mg/Fe] and [Ca/Fe] ratios relative to the 
field halo stars and the similarity in these abundance ratios with  
some of the more luminous dwarf spheroidal galaxies 
at metallicities [Fe/H]$<-2$ can be interpreted as 
star formation in Bo\"{o}tes I having lasted at least 
until Type Ia supernovae started 
to contribute to the chemical enrichment in this galaxy. 
}


   \maketitle
%

\section{Introduction}
\label{sec:intro}

Ultra-faint dwarf spheroidal galaxies (UFDs), recently discovered 
in the Local Group, provide us with 
an excellent laboratory for studying the formation of 
low-mass galaxies and production of heavy elements in their earliest stages 
in the Universe.
The faint Bo\"{o}tes I dwarf spheroidal galaxy (dSph) is one of the most 
interesting objects of its kind as a candidate for first galaxies 
and/or as building blocks of the Milky Way (MW). 
It was first discovered 
with the help of photometric data obtained by the Sloan Digital 
Sky Survey \citep[SDSS;][]{york00} as one of the faintest 
galaxies known to date \citep{belokurov06}.  
Subsequent spectroscopy to measure line-of-sight velocities 
of candidate member stars suggests that 
Bo\"{o}tes I is a dark-matter-dominated system with a total mass on 
the order of $10^{7} M_{\odot}$ \citep{munoz06,martin07}. 
\citet{koposov11} has developed a reduction technique to derive 
more accurate radial velocities, and they suggest that the 
velocity dispersion of this system is represented better by 
two components with 2.4 km s$^{-1}$ for a "cold" 
component and $\sim 9$ km s$^{-1}$ for a "hot" one, 
which could be interpreted as the result of velocity anisotropy in this galaxy.

Detailed analyses of a color-magnitude diagram (CMD) for Bo\"{o}tes I  
suggests that it consists of a very old stellar population with age  
$>10$ Gyr, which is similar to the most metal-poor globular 
clusters in the MW \citep{okamoto12}.
Chemical abundance analyses based on lower resolution spectroscopy 
report very low metallicities for Bo\"{o}tes I stars at [Fe/H]$\sim -2.5$ 
on average with a large spread of $\Delta$[Fe/H]$\gtrsim 1.0$ dex 
\citep{norris08,lai11}.

Detailed chemical abundances for Bo\"{o}tes I allow us to 
further address whether possible progenitors of the metal-poor 
component of the MW are similar to  Bo\"{o}tes I  and whether this 
galaxy is indeed an example of the first galaxies formed in the 
early Universe. However, these questions have not 
yet been clearly answered because of the discrepancy in abundance results 
among previous studies. 
Chemical abundances based on high-resolution 
spectroscopy for Bo\"{o}tes I stars have previously been obtained
 by \citet[][hereafter F09]{feltzing09}, \citet{norris10a}, 
and \citet[][hereafter G13]{gilmore13}. 
\citet{feltzing09} analyzed seven stars with $-3\lesssim$[Fe/H]$\lesssim-2$ 
based on the high-resolution spectra using an LTE abundance 
analysis code that takes the sphericity of stellar atmospheres into account.  
 They report the presence of a star with an anomalous 
[Mg/Ca] ratio, Boo-127, and interpret it as a 
signature of nucleosynthesis yields
of individual supernove (SNe). 

\citet{gilmore13} carried out the double-blind analysis by two groups 
of the co-authors, Norris \& Yong (NY) and Geisler \& Monaco (GM), for 
 six stars in Bo\"{o}tes I with $-3.7<$[Fe/H]$<-1.9$.  
They find that the chemical abundance pattern for the Bo\"{o}tes I 
stars are generally 
similar to those observed in field halo stars with small scatter.  
However, in contrast to what is generally known for the field halo stars with 
the similar metallicities, 
a signature of decreasing [$\alpha$/Fe] ratios with increasing [Fe/H] 
is also suggested.
They observed Boo-127 in common with F09 but 
did not confirm the [Mg/Ca] anomaly of this object.
Although these data provide important implications for the early 
chemical enrichment of this galaxy, they provide the contrasting pictures, 
namely, the former study reports the presence of an anomalous star, 
which may suggest that the chemical evolution in Bo\"{o}tes I proceeded in an 
inhomogeneous manner, while the latter does not support 
the presence of this inhomogeneity. 

To obtain a consistent picture of the early chemical 
evolution of Bo\"{o}tes I, we present an independent 
analysis of six Bo\"{o}tes I
stars with $-3.2<$[Fe/H]$<-2.0$ based on high-resolution spectroscopic 
data obtained with the Subaru/High Dispersion Spectrograph (HDS). 
Five stars and three stars in our sample have been analyzed in common 
with F09 and G13, respectively. 

This paper is organized as follows. In Section \ref{sec:obs}, we describe 
our sample, the method of observation, and 
equivalent width (EW) measurements. Section 
\ref{sec:analysis} presents the method for the 
stellar parameter and abundance 
estimates for the sample stars. In Section \ref{sec:results} we 
report the results of our abundance analysis and their comparison 
with previous results in the literature. Section \ref{sec:discussion} 
first presents the abundance comparison of the present Bo\"{o}tes I 
sample with the field MW halo stars and with other dSphs. 
Then, possible implications of the chemical evolution of 
Bo\"{o}tes I are discussed. 
Finally, Section \ref{sec:conclude} concludes the paper.


\section{Observation and measurements \label{sec:obs}}

\subsection{Sample and observation}

The sample of six bright red giant branch stars in Bo\"{o}tes I 
was selected from \citet{norris08} to cover a
wide metallicity range ($-3\lesssim$[Fe/H]$\lesssim-2$). 
The high-resolution spectra of
six bright red giants were obtained with the Subaru Telescope High
Dispersion Spectrograph \citep[HDS; ][]{noguchi02} on 16-17 May
2009 and 16 May 2010. The objects are listed in Table \ref{tab:obj} with 
observation details. 
We applied CCD on-chip binning (2$\times$2 pixels), 
resulting in a resolving power of 
$R=\lambda / \delta \lambda = 40,000$. Our spectral ranges 
cover 4030--6800~{\AA}. 
Spectra of two comparison stars in the Galactic halo (HD~216143 and
HD~85773) were obtained with similar setups of the instrument in our
previous run \citep{aoki09}.

A standard data reduction was made with IRAF echelle
package. Cosmic-ray hits were removed by the method described in
\citet{aoki05}. The observations were carried out mostly 
during dark nights, and the
sky background was not significant. One exposure of Boo--121 on 16 May 2010 was
exceptionally affected by the Moon. We thus subtracted the sky background 
estimated from the slit image around the object. The spectra obtained
from individual exposures were combined to obtain final spectra.

Radial velocities of the sample stars are measured from 25-50 Fe I
lines for each object. 
The results are given in Table \ref{tab:obj} ($v_{\rm helio}$).  
The measured values agree with those derived by G13 
within 1.5 km s$^{-1}$ for the three stars observed in common. 

\subsection{Equivalent width measurement}

The EWs of absorption lines were measured by Gaussian
fitting. The atomic line list used in \citet{ishigaki12, ishigaki13} was
adopted. The errors in the EW measurements estimated from the 
formula presented in \citet{cayrel88} range from $\sim 5$ to 12 m {\AA}, 
depending on the signal-to-noise ratios ($\sim 10$ per pixel 
at $\sim 4500$ {\AA} 
and $\sim 30$ at $\sim 6000$ {\AA}) of the continuum.
 
Although Mg is a key element in our work, the {\it gf} values 
could be uncertain because of
the lack of laboratory measurements for most of the Mg I lines
\citep{aldenius07}. 
We inspected the {\it gf} values of Mg I lines adopted
in previous studies for Bo\"{o}tes I stars F09 and G13 and summarize them 
in Table \ref{tab:mglines} (the values of F09 was provided by
Feltzing, private communication). The last column gives the 
$\log gf$ values adopted in this work. The Mg I lines at 5172.68, 5183.60, 
and 5711.09 {\AA} were not used in this work because EWs 
of these lines for our sample stars were too strong or too weak to be 
useful in the abundance analysis.
As can be seen from Table \ref{tab:mglines}, 
the $\log gf$ values for the 5528.40 {\AA} line, 
which were common to all of these studies, are different 
by up to $\sim0.3$ dex. 

 We confirmed that the Mg abundances derived from different lines 
using the $\log gf$ values 
adopted in this work (last column of Table \ref{tab:mglines})
 agree within $\sim$0.3 dex for the comparison stars, HD~216143 
and HD~85773, for each.

\begin{table*}
\caption{Objects and observations}
\label{tab:obj}
\centering
\begin{tabular}{lcccccc}
\hline\hline
Name & RA(J2000) & DEC(J2000) & $g$ & $(g-r)_{0}$ & $v_{\rm helio}$ (km s$^{-1}$) & HJD \\ 
\hline
Boo--094 & 14:00:31.50 & 14:34:03.6 &   17.50 &   0.87 &   95.0$\pm$0.7 &    2454968 \\
Boo--121 & 14:00:36.52 & 14:39:27.3 &   17.92 &   0.82 &  105.4$\pm$0.6 &    2454968 \\
Boo--911 & 14:00:01.07 & 14:36:51.5 &   17.97 &   0.80 &  102.2$\pm$0.6 &    2454969 \\
Boo--127 & 14:00:14.56 & 14:35:52.7 &   18.16 &   0.77 &   98.5$\pm$0.5 &    2455333 \\
Boo--009 & 13:59:48.81 & 14:19:42.9 &   17.93 &   0.81 &   97.1$\pm$0.4 &    2455333 \\
Boo--117 & 14:00:10.49 & 14:31:45.5 &   18.20 &   0.74 &   98.0$\pm$0.6 &    2455333 \\
\hline
\end{tabular}
\end{table*}

\begin{table}
\caption{The Mg linelists. }
\label{tab:mglines}
\centering
\begin{tabular}{lccccc}
\hline\hline
  &  &  F09\tablefootmark{a} & G13(NY)\tablefootmark{a} & G13(GM)\tablefootmark{a} & TW\tablefootmark{a} \\ 
$\lambda$ & $\chi$  & \multicolumn{4}{c}{$\log gf$}\\ 
 (\AA) & (eV) & \multicolumn{4}{c}{(dex)} \\ 
\hline
    4571.10 &    0.00 &  ... &  ... &  ... & $-$5.623\\ 
    4702.99 &    4.35 & $-$0.666 &  ... &  ... & $-$0.440\\ 
    4730.03 &    4.35 &  ... &  ... &  ... & $-$2.347\\ 
    5172.68 &    2.71 & $-$0.402 &  ... &  ... &  ...\\ 
    5183.60 &    2.72 & $-$0.180 &  ... &  ... &  ...\\ 
    5528.40 &    4.35 & $-$0.620 & $-$0.340 & $-$0.620 & $-$0.498\\ 
    5711.09 &    4.35 &  ... &  ... & $-$1.833 &  ...\\ 
\hline
\end{tabular}
\tablefoot{ 
\tablefoottext{a}{The $\log gf$ values for the Mg lines
 adopted in F09, G13, and this work (TW). The 
two independent analyses carried out by G13 are referred to as 
"NY" and "GM", respectively (see Section \ref{sec:intro}).}
}
\end{table}

\subsection{Equivalent width comparisons{\label{sec:ew}}}

\begin{figure*}
\centering
\includegraphics[width=14.0cm]{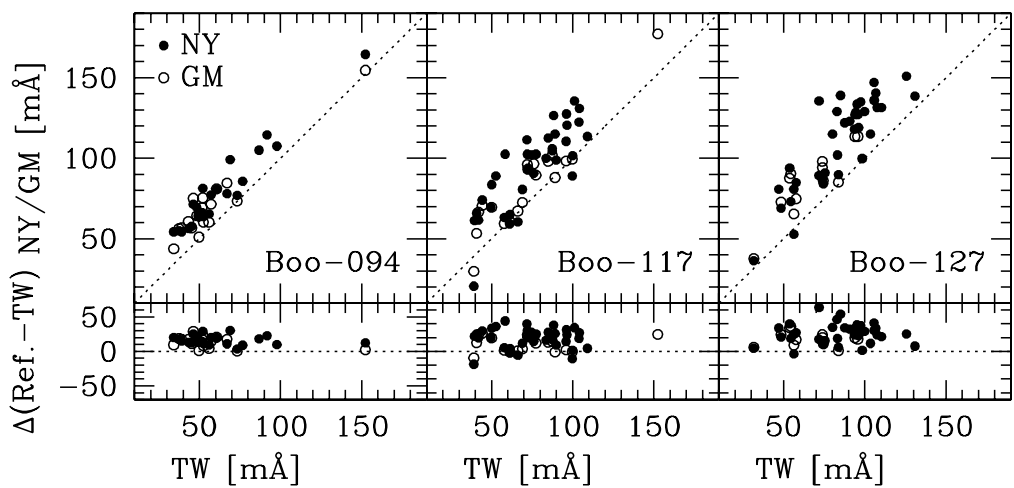}
\caption{Comparison of the measured equivalent widths (EWs) in 
this work (TW) and 
in the two analyses, "NY" (closed circles) and "GM" (open circles), 
in \citet{gilmore13}. }
\label{fig:comp_ew}
\end{figure*}

Figure \ref{fig:comp_ew} presents the comparison between the EWs 
measured in this 
work and those in G13 for the three stars analyzed in common. 
For all of these stars, the EWs measured in this work 
are systematically lower than the values obtained by G13. 
The mean differences between the EW in this work and in 
the analysis of "NY" in G13 are 16-26 m{\AA}, while 
those between in this work 
and in the other independent analysis, "GM", in G13 are 
10-18 m{\AA}, which are modestly larger than the errors in the EWs 
measurement ($\sim 12$ m{\AA}). 

\begin{figure}
\centering
\includegraphics[width=8.5cm]{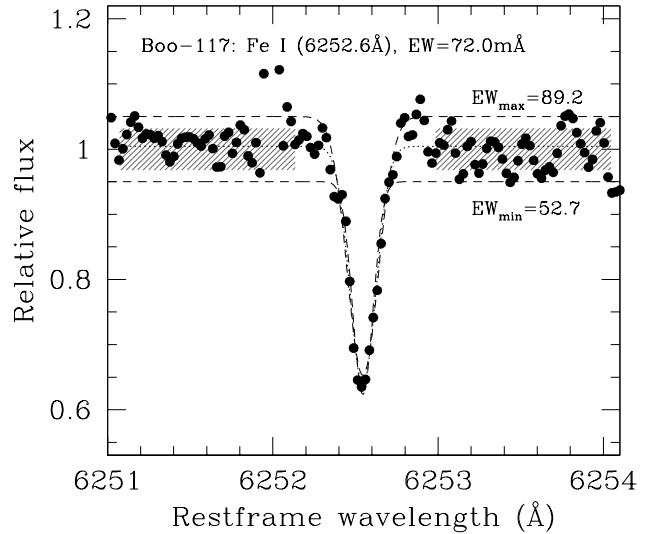}
\caption{EW measurements for one of the 
Fe {\small I} lines by adopting the different 
continuum levels. The dotted line shows the best-fit Gaussian, where 
continuum level is estimated from the shaded wavelengths region. 
The dashed lines show the Gaussian fit by fixing the 
continuum levels to be $\pm 0.05$ of the best-fit value. }
\label{fig:ew_cont}
\end{figure}

To test whether the uncertainty in the continuum 
placement could be the reason for the discrepancy, 
we examined the change in the measured EWs by adopting different continuum 
levels for one of the Fe {\small I} line as shown in
 Figure \ref{fig:ew_cont}. 
For this line, the discrepancy 
in the measured EWs among the three analyses is particularly 
large (72.0 m{\AA}, 96.2 m{\AA} and 102.5 m{\AA} in this work, 
GM, and NY, respectively). 
If the continuum level of the normalized spectrum 
is changed by $\pm 0.05$, which is slightly more than 
the scatter in the continuum around this line (widths in the 
shaded regions), 
the measured EW varies in the range $52.7-89.2$ m{\AA}.
This result illustrates that the variation in the continuum level 
partly explains the EW difference among the three analyses. 
However, as can be seen in Figure \ref{fig:ew_cont}, the error in 
the continuum level that is as large as $\pm 0.05$ is unlikely 
for this line, so it is difficult to completely illuminate the discrepancy.  
In this paper, we use the measured EWs without any correction. 
Instead, we examine the 
effects of this offset in derived abundances 
in Section \ref{sec:abuerror}.

\section{Abundance analysis \label{sec:analysis}}
\subsection{Stellar atmospheric parameters}
Effective temperatures ({\teff}) were estimated based on $griz$
color indices using the temperature scale of 
Castelli\footnote{http://wwwuser.oat.ts.astro.it/castelli/colors/sloan.html}.  
The $griz$ photometry was adopted from the SDSS DR9 data set. 
The Galactic extinction in the $griz$ magnitudes 
toward individual stars was corrected 
based on the revised extinction measurement by \citet{schlafly11} via 
the NASA/IPAC Extragalactic Database\footnote{http://ned.ipac.caltech.edu/forms/calculator.html}. The extinction in the $g$-band varies from 0.052 to 0.060 
among our six sample stars.  These values 
are translated into {\ebv} in the \citet{schlegel98} from 0.016 to 0.018 
\citep{schlafly11}, 
which are consistent with the previously adopted 
value of {\ebv}$=0.02$ \citep{norris08}. 

The surface gravity ($\log g$) for the sample stars are then estimated 
by comparing the adopted temperature with a 12 Gyr Yonsei-Yale (YY) isochrone 
\citep{demarque04} by assuming that 
the stars lie on the red giant branch  
of the system. In the above procedure for obtaining {\teff} and {\logg}, 
the [Fe/H] values are needed as input. We adopt the initial 
guess of the [Fe/H] values from \citet{norris10b}. 

The microturbulent velocity ($\xi$) is estimated by 
requiring that the Fe abundances from individual Fe {\small I} lines 
do not depend on their measured EWs. In this step, 
we have used the Fe {\small I} lines having a reduced EW 
$\log (EW/\lambda)<-4.7$. 

After the standard 1D-LTE abundance analysis is performed, 
the procedures are iterated until a consistent 
set of parameters ({\teff}, $\log g$, $\xi$, and [Fe/H]) is obtained. 
The resulting atmospheric parameters are summarized in 
Table \ref{tab:stpm}. 

The adopted {\teff} and {\logg} in this work and in 
the two independent analyses (``NY'' and "GM") presented 
by G13 agree within 100 K and 
0.2 dex, respectively,  for the three stars analyzed in common.  
The $\xi$ values are different between the 
two analyses in G13 by 0.5-1.3 km s$^{-1}$. 
The relatively high $\xi$ value for Boo--094 reported by the 
"NY" analysis is in excellent agreement with ours, while 
the $\xi$ values for Boo--117 and Boo--127 are more 
consistent with those obtained by the "GM" analysis. 

\begin{table}
\caption{Atmospheric parameters}
\label{tab:stpm}
\centering
\begin{tabular}{lcccc}
\hline\hline
Object & $T_{\rm eff}$ & $\log g$ & $\xi$ & [Fe/H] \\ 
 & (K) & (dex) & (km s$^{-1}$) & (dex) \\ 
\hline
   Boo--094 & 4500 &    0.8 &    3.4 &   $-$3.2 \\ 
   Boo--121 & 4500 &    0.8 &    2.1 &   $-$2.5 \\ 
   Boo--911 & 4500 &    0.9 &    1.7 &   $-$2.2 \\ 
   Boo--127 & 4750 &    1.6 &    1.6 &   $-$1.9 \\ 
   Boo--009 & 4750 &    1.4 &    2.5 &   $-$2.7 \\ 
   Boo--117 & 4750 &    1.5 &    1.9 &   $-$2.2 \\ 
  HD216143 & 4529 &    1.3 &    1.9 &   $-$2.1 \\ 
   HD85773 & 4366 &    1.0 &    2.1 &   $-$2.4 \\ 
\hline
\end{tabular}
\end{table}

\subsection{Abundances}

Abundances of individual elements are determined by standard
analyses using the measured EWs. 
We restrict the analysis to the absorption lines with 
$\log (EW/\lambda)< -4.7$ to minimize the errors in derived 
abundances. For Na {\small I} lines, we relax 
this restriction to $\log (EW/\lambda)\le -4.5$, since only strong 
resonance lines are detected for this species. 
Similarly, only strong resonance Sr {\small II} lines are available for 
the abundance estimate of Sr.  A spectral synthesis is applied 
to estimate Sr abundances as described in detail in Section \ref{sec:nc}. 
The hyperfine splitting effect is
included in the analysis of Ba lines \citep{mcwilliam98}, assuming the
r-process isotope ratios. 
The EWs and abundances estimated from individual 
lines are summarized in Table \ref{tab:ew}.

Carbon abundances are determined from the comparisons of synthetic
spectra to the observed spectrum of the CH molecular band at around
4323~{\AA}. For Boo--094, Boo--121, and Boo--009, only an 
upper limit of carbon abundance is
estimated. Because the signal-to-noise ratios are low for the wavelength 
range of the G-band, the continuum level is not determined well,
which produces the abundance uncertainty of $0.1-0.3$ dex.

\subsection{Abundance errors}
\label{sec:abuerror}

The errors due to the line-to-line scatter ($\sigma_{\rm line}$), 
which are calculated 
as the standard deviation of the abundances from 
individual lines divided by the square root of the 
number of lines, are summarized in Table \ref{tab:abund}. 
If only one line is available for the abundance 
estimate, $\sigma_{\rm line}$ is assumed to be equal to 
the standard deviation for the Fe {\small I} lines, which 
are typically $\sim 0.2$ dex in the present work. 

The abundance errors due to uncertainties in {\teff}, {\logg}, and 
$\xi$ are examined by calculating the 
abundance differences when changing these parameters
by $\pm 100$K, $\pm 0.3$ 
dex, and $\pm 0.3$ km s$^{-1}$, respectively. Results 
of this exercise are summarized in Table \ref{tab:abund}.
The error bars 
in Figures \ref{fig:xfe} and \ref{fig:nc}, and $\sigma_{\rm tot}$ 
in Table \ref{tab:abund} correspond to a 
quadratic sum of the 
contribution from the uncertainty in the three 
atmospheric parameters and line-to-line 
scatter in the abundances. 

\begin{figure}
\centering
\includegraphics[width=8.5cm]{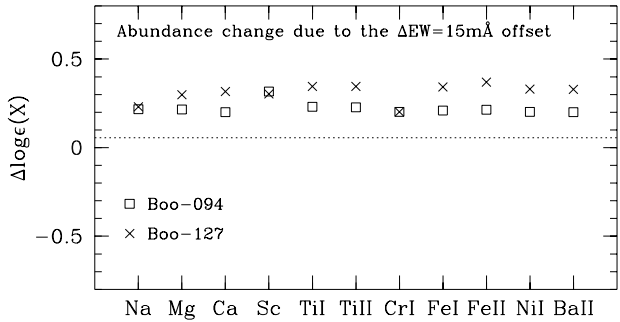}
\caption{ Changes in abundances when all of the EWs in this work are increased 
by 15m{\AA} to partly correct for the EWs offset from \citet{gilmore13}. }
\label{fig:deqw}
\end{figure}

The systematic offset in 
the EWs from those in \citet{gilmore13} mentioned in 
Section \ref{sec:ew} are an additional source of uncertainty in 
measuring absolute abundances in this work. Figure \ref{fig:deqw} shows 
changes in $\log \epsilon (X)$ abundances when the EWs are 
increased by 15m{\AA} to reduce the systematic offset from 
the \citet{gilmore13} analyses for the most metal-poor 
(Boo--094) and metal-rich (Boo--127) stars in our sample. 
As can be seen, after the EW offset is applied, the abundances  
are systematically increased by $\sim 0.15-0.30$ dex. 
On the other hand, the amount of correction is mostly similar 
for all elements, and thus difference in the abundance ratios relative to 
iron ([X/Fe]) partly cancel out. 
Although the absolute abundances could be affected by the
systematic errors in EW measurements, the abundance ratios ([X/Fe]) are 
consistent with those in \citet{gilmore13}. 
 
For the Sr abundances estimated by the spectral syntheses, 
uncertainty in the continuum level ($1\sigma \sim 0.1$) 
causes an error as large as $\pm 1.0$ dex.  as detailed in Section \ref{sec:nc}.

\section{Abundance results}
\label{sec:results}

Our abundance results are summarized in Table \ref{tab:abund}. 
Relative abundances ([X/Fe]) were determined adopting the 
solar abundances from \citet{asplund09}.
 Figures \ref{fig:xfe} and \ref{fig:nc} plot the 
relative abundances of 
C, Na, Mg, Ca, Sc, Ti {\small I}, Ti {\small II}, Cr, Ni, Zn, Sr, and Ba 
as a function of [Fe/H] for the sample of six Bo\"{o}tes I 
stars (filled circles) and the two 
comparison field halo stars (open circles). Triangles 
represent the abundance results from the "GM" analysis in 
G13, whose systematic offset in the EWs from those in this work 
is smaller than that in the ``NY" analysis.

\subsection{Comparison with \citet{feltzing09}}

Five objects in the present sample have also been studied by 
F09, who determined Fe, Mg, Ca and Ba abundances for 
seven objects in Bo\"{o}tes I. 
The [Fe/H] of the sample stars determined by the present work 
agree well with the results of F09 within $\sim 0.1$ dex, 
except for the lowest metallicity star Boo-094 in our sample, for which 
the estimated [Fe/H] is $-3.4$ dex in our study but $-2.9$ in 
F09. This difference can be partly attributed to the higher $\log g$ 
 and larger $\xi$ adopted in our analysis. 
 
The abundances of Mg and Ca also marginally agree with those estimated by 
F09 within 0.30 dex, 
except for Boo-127. For this star, F09 
report an exceptionally high [Mg/Ca] ratio ([Mg/Ca]$>0.6$), while this 
work has obtained [Mg/Ca]=0.08. 
The estimated $\log \epsilon$(Mg) in this work is $\sim 0.5$ dex lower 
than that reported by F09. 
Several factors seem to be responsible for the abundance difference, 
including (1) a higher $\log gf$ value for Mg {\small I} 5528 {\AA} 
(see Table \ref{tab:mglines}), 
(2) higher $\log g$, and (3) higher $\xi$ in this work than those 
adopted in F09. While the error in the EWs by up to 
$\sim 10-15$ m{\AA} may also contribute to the discrepancy, 
it does not totally explain the large difference. 

\begin{figure}
\centering
\includegraphics[width=8.5cm]{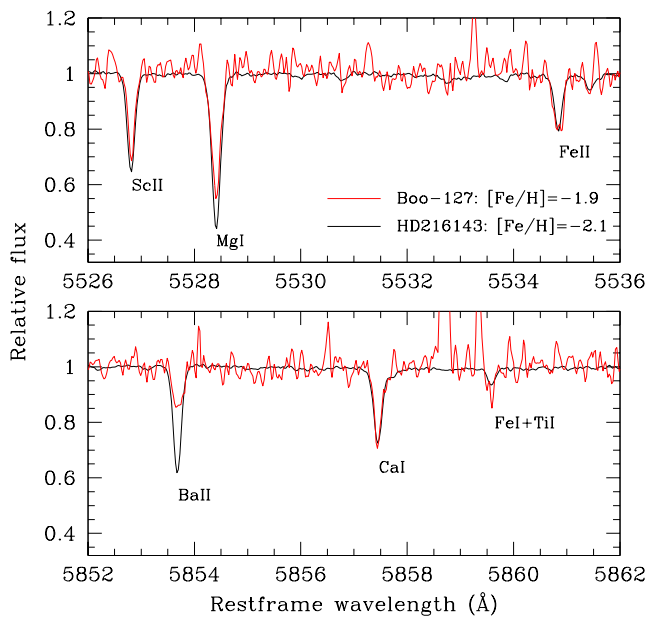}
\caption{Spectra of Boo--127(red) and the comparison star, 
HD 216143 (black), for the 
wavelength regions that contain Mg {\small I} and Ca {\small I} lines.}
\label{fig:plotdiff}
\end{figure}

To illuminate the uncertainty in the $\log gf$ values, 
we also performed a simple differential analysis to 
obtain [Mg/Ca] for Boo--127 with respect to 
one of the comparison stars, HD 217143. 
Since the $T_{\rm eff}$, $\log g$, and [Fe/H] values are 
similar between the two stars, a direct comparison of EWs for 
the Mg {\small I} and Ca {\small I} lines provides the [Mg/Ca] estimate. 
Figure \ref{fig:plotdiff} shows the wavelength regions 
that contain the Mg {\small I} and Ca {\small I}  lines for Boo-127 
and HD 216143. While both stars show Ca {\small I} line 
with similar strengths, the Boo-127 shows slightly weaker 
Mg {\small I} line than 
HD 216143, which does not support the anomalously high [Mg/Ca] 
in Boo-127 compared to the reference star. 
More quantitatively, The relative abundance $\log(A/A_{\rm ref})$ 
is expressed as

\begin{equation}
\log\left(\frac{A}{A_{\rm ref}}\right)=\log\left(\frac{EW}{EW_{\rm ref}}\right)-\log\left(\frac{\kappa_{\nu}}{\kappa_{\nu,{\rm ref}}}\right)-\Delta\theta_{\rm ex}\chi
\end{equation}
where $\kappa_{\nu}$ and $\kappa_{\nu, {\rm ref}}$ are continuum opacity, 
$\Delta\theta_{\rm ex}$ ($\theta=5040/T$) is the difference 
in an excitation temperature, and $\chi$ is the excitation potential 
of the line \citep{gray05}. If we assume that the 
$\kappa_{\nu}$ term is negligible and $\Delta\theta_{\rm ex}$ is constant 
and equal to the one computed from the effective temperature
($\Delta\theta_{\rm ex} = \Delta\theta_{\rm ex, eff}$) 
for all relevant optical depths,
the relative abundance 
can be obtained from the ratios of the EWs \citep{smith82}. 
This results in the [Mg/Ca] ratio for Boo--127 being $\sim 0.04$ dex 
lower than that of HD 216143, which is, again, inconsistent 
with the anomalously high [Mg/Ca] ratio. 
The low, but not extremely low, Ba abundances ([Ba/Fe]$\sim
-0.6$) of Boo-121 and Boo-911 found by F09 are 
confirmed by our measurement.
The $\log \epsilon$(Ba) in the two studies agree within 0.3 dex.

\subsection{Comparison with \citet{gilmore13}} 

Comparisons of the relative abundances in this work and 
in the "GM" analysis of G13 are presented 
in Figure \ref{fig:xfe} (triangles that are connected with the data points 
in this work) for the three stars analyzed in common. 
Within the quoted error bars, our abundance results generally 
agree with those obtained by G13 with a few  
exceptions. One relatively large discrepancy is found in 
the [Fe/H] value for Boo-094, the most metal-poor 
star in our sample, which is $\sim 0.2$ dex lower in 
this work than obtained by G13. 
The smaller EWs (see Section \ref{sec:ew}) by $\sim 10$m{\AA} 
and larger $\xi$ adopted in this work 
than in G13 are probably responsible for the 
discrepancy.

\subsection{Carbon} 
The carbon abundance is estimated for three of the six sample stars, 
while only an upper limit is obtained for the other three objects. 
For the six sample stars, extremely 
carbon-rich objects are not found. These sample stars all show 
[C/H]$<-2.7$ and [C/Fe]$<0.2$. For the three objects with 
measured carbon abundances, the average [C/Fe] is $-0.7$ dex with 
small scatter. 
These values are close to the lowest bounds of the [C/Fe] distributions 
obtained by \citet{lai11} and \citet{norris10b} for this galaxy. 
This [C/Fe] value is comparable 
to what is seen in field halo stars 
having similar luminosity \citep{spite05,norris10b}.

\subsection{Sodium}

For the present Bo\"{o}tes I sample stars, sodium abundances are 
only estimated using the strong Na {\small I} resonance 
lines at 5889/5895 {\AA}. 
These lines are known to suffer from strong
non-LTE (NLTE) effects such that LTE abundances tend to 
be larger than corresponding NLTE values \citep{takeda03}. 
The NLTE correction to Na abundances ranges from $-0.1$ dex to 
$-0.5$ dex, depending on the atmospheric parameters 
of stars \citep{takeda03}.
Without any NLTE correction, the average 
[Na/Fe] value for the six Bo\"{o}tes I stars is $\sim -0.1$ dex with 
small scatter. This value is similar to the uncorrected 
 [Na/Fe] values obtained for field halo stars with comparable 
metallicities.

\subsection{$\alpha$/Fe ratios}

The [Mg/Fe] ratios in our sample stars are similar to or slightly lower than  
those obtained for the field halo stars. 
A hint of a modest decreasing trend of the [Mg/Fe] with increasing [Fe/H] is 
seen, as was also reported in G13. 
The [Ca/Fe] ratios are also similar to or slightly lower than the value 
for the field halo stars.
 Excluding the lowest metallicity star, 
a scatter in the [Ca/Fe] ratios is remarkably 
small ($\sim 0.1$ dex).
The [Ti {\small I}/Fe {\small I}] and 
[Ti {\small II}/Fe {\small II}] ratios differ by up to $\sim 0.3$ dex in our 
Bo\"{o}tes I sample stars, as reported for the field halo stars 
\citep[e.g.,][]{lai08}.  
The Bo\"{o}tes I stars tend to 
show slightly lower [Ti {\small I}/Fe {small I}]  abundance ratios 
than the halo stars, while 
 the [Ti {\small II}/Fe {\small II}] ratios for these samples are similar.

\subsection{Iron-peak elements} 

Scandium abundances are estimated from a few to several 
Sc {\small II} lines in our sample stars. The mean abundances 
are similar to or slightly lower than those reported for 
the field halo stars in [Fe/H]$>-2.5$. 

Chromium abundances are obtained from a few 
Cr {\small I} lines for the Bo\"{o}tes I stars. The [Cr/Fe] ratios 
are similar to 
those reported for the field halo stars for two of the  
most metal-rich stars, while other stars show lower [Cr/Fe] ratios. 
However, since only one or two lines with low-signal-to-noise ratio 
are used in the abundance estimate, the significantly lower 
[Cr/Fe] ratios for these stars require further confirmation.

In the four higher metallicity stars in our sample, 
the [Ni/Fe] ratios are similar to those obtained for 
the field halo stars, except for one object (Boo--121).
Two Ni {\small I} lines used to obtain Ni abundance 
in this stars both suggest a subsolar [Ni/Fe] ratio.

\begin{figure*}
   \centering
   \includegraphics[width=14.0cm]{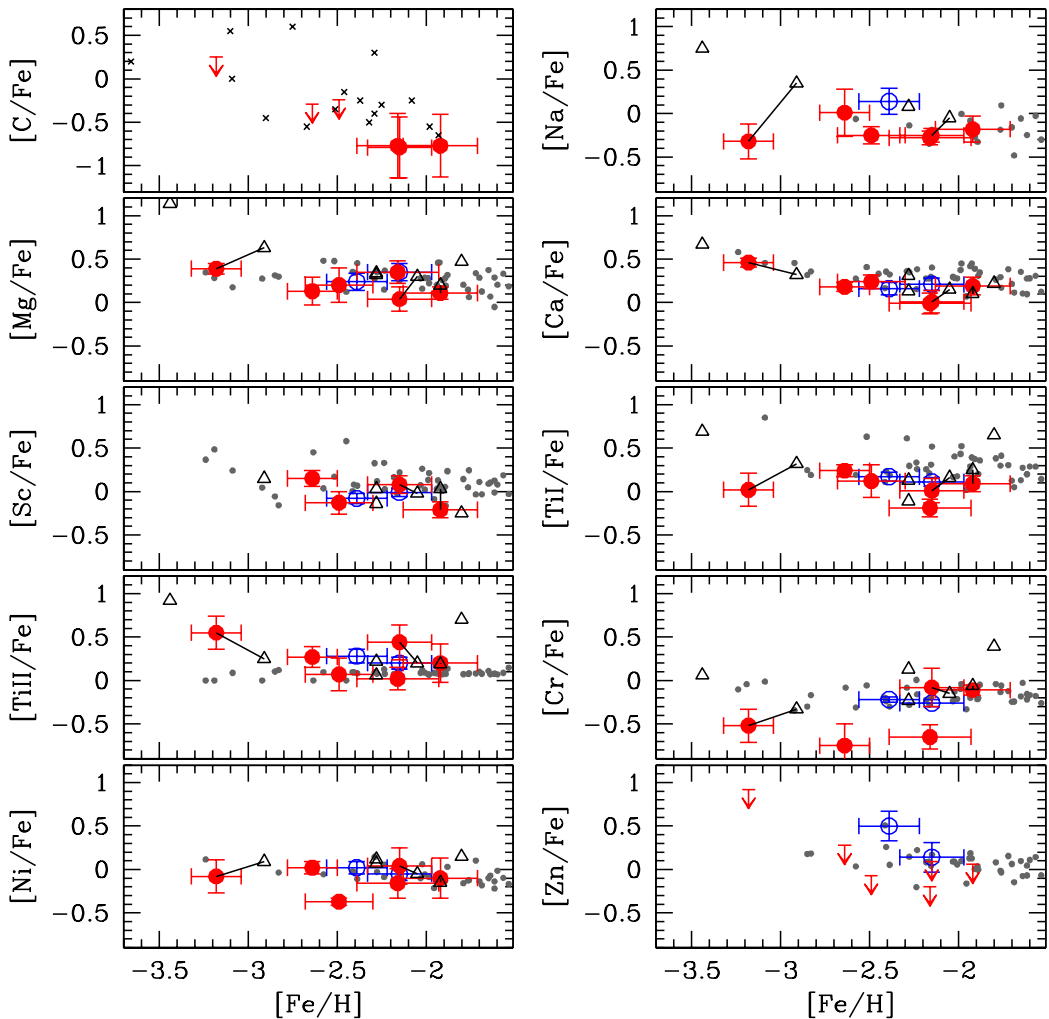}
   \caption{[X/Fe] ratios plotted against [Fe/H] for the sample of 
six Bo\"{o}tes I stars (red circles) and the two comparison 
field halo stars (blue circles). Black triangles represent the 
abundance results from the "GM" analysis in G13. 
Small gray circles present the abundances of field halo stars from 
\citet{ishigaki12,ishigaki13}. In the plot for [C/Fe]-[Fe/H], results from 
\citet{norris10b} for Bo\"{o}tes I stars are shown in crosses.}
   \label{fig:xfe}
   \end{figure*}

Since EWs of Zn {\small I} lines are not reliably measured 
for our sample stars, we used spectral synthesis to 
estimate upper limits of the Zn abundances.
The results are shown by bars with downward arrows 
in the bottom right hand panel of Figure \ref{fig:xfe}. 
The upper limits of the [Zn/Fe] ratios are similar to 
or lower than the field halo samples in the metallicity range 
[Fe/H]$>-2.5$.

\subsection{Neutron-capture elements \label{sec:nc}}  

\begin{figure}
   \centering
   \includegraphics[width=8.5cm]{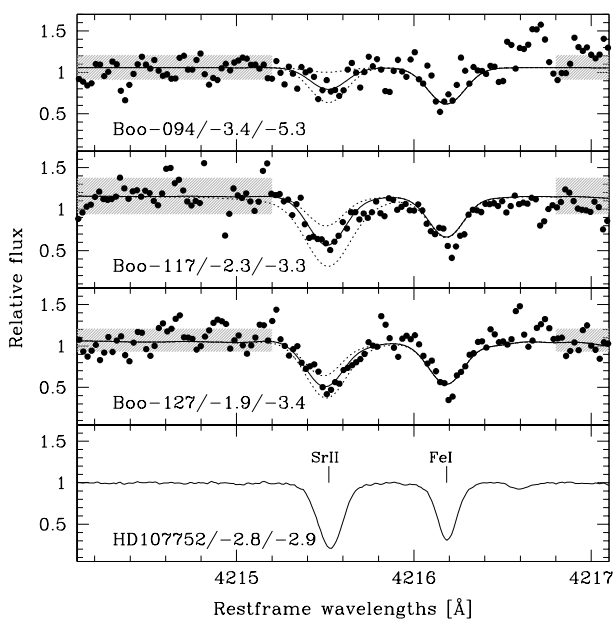}
      \caption{Observed (black dots) and the best-fit synthetic 
(solid line) 
spectra of the three Bo\"{o}tes I stars around the Sr II 
4215.5 {\AA} line. The obtained [Fe/H] and [Sr/H] values are 
indicated in the bottom-left corner of the each panel ("Name/[Fe/H]/[Sr/H]"). 
The dotted lines show the synthetic spectra fitted after changing 
the continuum levels by $\pm 1\sigma_{\rm STDV}$ (in each panel, the 
continuum levels have been shifted to match the best-fit value). 
For comparison, the spectrum of one halo giant, HD 107752, is shown 
in the bottom panel.}
         \label{fig:plt_sr}
   \end{figure}

The strontium abundances are obtained by fitting synthetic spectra
to the observed ones for the Sr {\small II} resonance line at 4215.5 {\AA}. 
The $\log gf$ value for this line is taken from NIST database. 
Figure \ref{fig:plt_sr} shows the resulting 
synthetic and the observed 
spectra around the Sr {\small II} line for the Bo\"{o}tes I stars for which 
the Sr abundances are reasonably obtained. The other three 
sample stars are excluded in the analysis since the spectral region 
is too noisy to estimate Sr abundances.

In the fitting procedure, we redefine the continuum level using 
the wavelength regions 4214.10--4215.25 {\AA} and 4216.80--4217.10 {\AA}. 
To estimate abundance errors due to the uncertainty in the 
continuum level, we performed the fitting changing the continuum 
level of the synthetic spectrum by $\pm 1\sigma_{\rm STDV}$, 
where $\sigma_{\rm STDV}$ is a standard deviation in the observed 
continuum level. 
The error estimated in this procedure is
 $\sim 0.7-1.5$ dex, as summarized in Table \ref{tab:abund}. 

For comparison, the bottom panel of Figure \ref{fig:plt_sr} 
shows the observed spectrum of a field halo giant, HD~107752, 
which has atmospheric parameters similar to those of the present sample 
({\teff}$=4820$ K, {\logg}$=1.6$, $\xi=1.9$ {\kms}, and [Fe/H]=$-2.8$) 
\citep{ishigaki13}. While the Sr {\small II} line of the HD~107752 
appears to be stronger than a nearby Fe {\small I} line, those in the three 
Bo\"{o}tes I stars seem to have comparable strength as the Fe {\small I} lines. 

\citet{andrievsky11} studied the NLTE effects on Sr abundance 
determination for extremely metal-poor stars and report that 
the NLTE correction for the Sr {\small II} 4215.5 
{\AA} line is at most $\sim$0.2 dex for stars
having atmospheric parameters similar to the present 
Bo\"{o}tes I sample. The correction is smaller than 
the other sources of uncertainty in the present analysis, 
so that is not included in our abundance results.

\begin{figure}
   \centering
   \includegraphics[width=8.5cm]{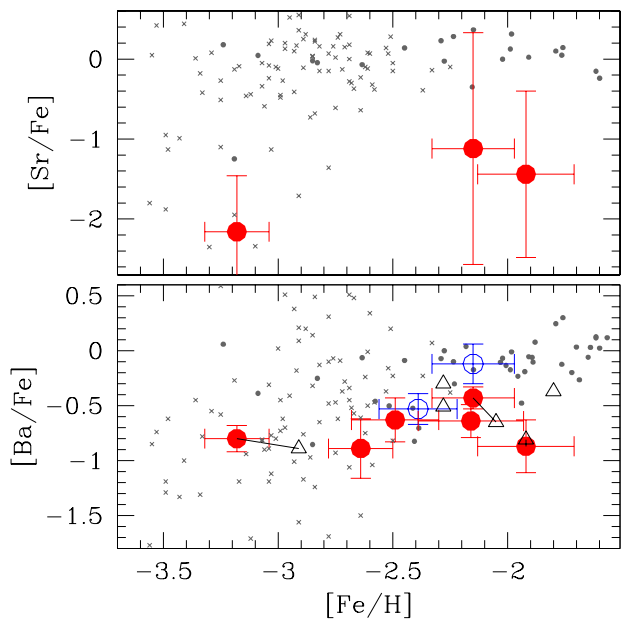}
   \caption{Same as Figure \ref{fig:xfe} but for Sr and Ba. Abundance data 
extracted from the SAGA database \citep{suda08} (RGB, 
$R>30000$ with no flags of "C-rich" nor ``CEMP") 
are additionally plotted with gray crosses to supplement the MW halo data. }
   \label{fig:nc}
   \end{figure}

The top panel of Figure \ref{fig:nc} shows resulting [Sr/Fe] values 
plotted against [Fe/H] with the error bars. 
The crosses and small filled circles represent 
the abundances from literature extracted 
from SAGA database and those from \citet{ishigaki13}, respectively. 
 As can be seen, even though the errors are very large, the [Sr/Fe] ratios of 
the three Bo\"{o}tes I stars are significantly  
lower than the solar value, which is not typical for 
the field halo stars at similar metallicities. 
Most of the field halo sample in literature show the solar 
value in [Fe/H]$<-2.5$ and increasing dispersion toward lower 
metallicity with [Sr/Fe]$\sim -2.0$ to $\sim 0.5$.
On the other hand, the Bo\"{o}tes I 
stars have [Sr/Fe]$\sim -1$ in [Fe/H]$>-2.5$, and the most metal-poor 
star in our sample show even lower [Sr/Fe] of $\sim -2$.

The bottom panel of Figure \ref{fig:nc} plots the [Ba/Fe] ratios against 
[Fe/H]. The [Ba/Fe] of Bo\"{o}tes I stars 
are subsolar, generally confirming the result of 
G13. The [Ba/Fe] ratios for these stars 
tend to distribute at the lower bound of 
 the [Ba/Fe] distribution for the field halo stars.  

One interesting feature is that Boo--127, 
which is the most metal-rich star in our sample, shows 
lower [Ba/Fe] than the field halo stars at similar 
metallicities ([Fe/H]$\sim -2$). 
As shown in Fig \ref{fig:plotdiff}, the Ba {\small II} 
line is significantly weaker in Boo--127 than 
seen in the comparison star with similar [Fe/H].

\section{Discussion \label{sec:discussion}}

\subsection{Comparison with the field MW halo}
\label{sec:compmw}

If the MW halo was assembled from smaller satellite galaxies similar 
to the surviving dSphs, 
as suggested by the standard cosmological model, we  
naively expect that chemical abundances in the MW field halo 
stars are similar to these dSphs. 
In this section, we address the question of whether the 
observed abundances in Bo\"{o}tes I stars are similar to those 
obtained for the field halo stars.    

\begin{figure}
   \centering
   \includegraphics[width=8.5cm]{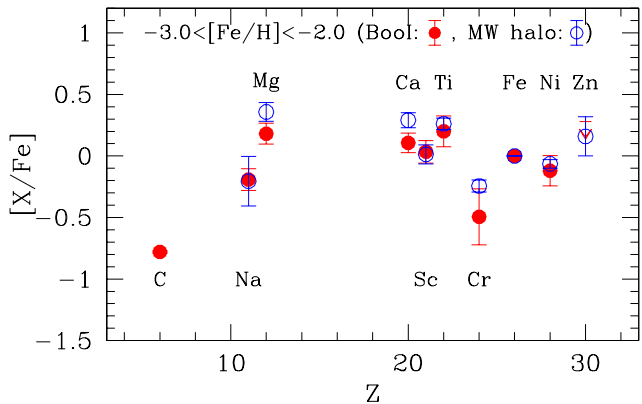}
   \caption{The abundance pattern of the Bo\"{o}tes I stars (filled circles) and the MW halo stars (open circles) with $-3.0<$[Fe/H]$<-2.0$. }
   \label{fig:pattern}
   \end{figure}

Figure \ref{fig:pattern} compares mean abundances of 
the Bo\"{o}tes I stars and field halo stars in the metallicity range of 
$-3.0<$[Fe/H]$<-2.0$. To minimize systematic errors in the abundance 
comparison, eight halo giants ({\teff}$<5000$K and {\logg}$<3.0$, including 
one of the comparison stars HD 85773) 
from \citet{ishigaki12,ishigaki13} are only considered
for the field halo sample. In their work, the adopted 1D-LTE 
abundance analysis technique and the line list are the same as 
in this work. 
Error bars in Figure \ref{fig:pattern} represent 
standard deviations around the mean values.

In comparison to the field halo stars, the Bo\"{o}tes I stars show 
slightly lower [Mg/Fe] and [Ca/Fe] ratios in the metallicity range of 
$-3.0<$[Fe/H]$<-2.0$. 
Kolmogorov-Smirnov (KS) tests suggest that the probability at which [X/Fe] 
ratios of the halo 
stars and the Bo\"{o}tes I stars can be drawn 
from the same distribution is less than $5$\% for Mg and Ca, 
 which marginally supports the abundance difference. 
On the other hand, the [Na/Fe] ratios for the Bo\"{o}tes I stars 
are indistinguishable from those of the halo stars. 

\begin{figure}
   \centering
   \includegraphics[width=8.5cm]{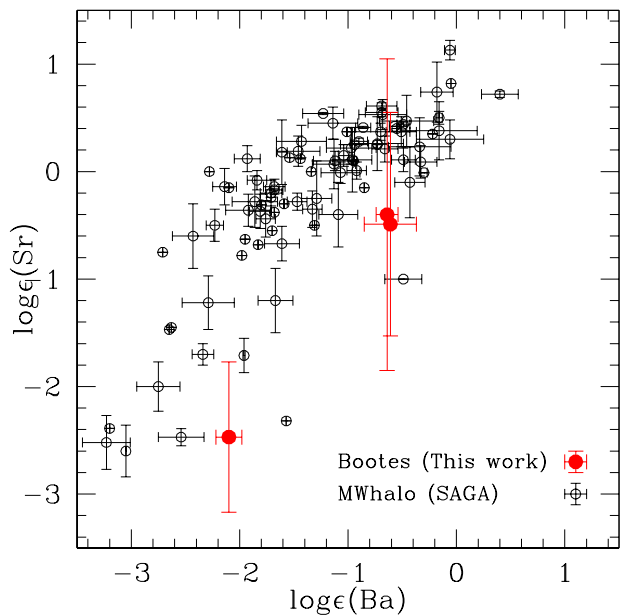}
   \caption{$\log\epsilon$(Ba) versus $\log\epsilon$(Sr) for the three Bo\"{o}tes I stars. For comparison, the abundance data for MW halo giants are extracted from SAGA database \citep{suda08} with the criteria $-3.5<$[Fe/H]$<-2.0$ and a spectral resolution $R>30000$ with no flags of ``C-rich" or ``CEMP''.  }
   \label{fig:ba_sr}
   \end{figure}

As mentioned in the previous sections, 
the Bo\"{o}tes I and the MW halo stars show
a hint of an abundance difference in Sr.  
Figure \ref{fig:ba_sr} shows $\log\epsilon$(Sr) plotted 
against $\log\epsilon$(Ba) for the Bo\"{o}tes I stars and 
MW halo giants. The data for the MW halo giants were 
extracted from the SAGA database \citep{suda08} with 
criteria, $-3.5<$[Fe/H]$<-2.0$ and a spectral 
resolution $R>30000$ with no flags of ``C-rich" or ``CEMP''.
As can be seen, most of the halo stars show Sr/Ba$>1$, 
while the Bo\"{o}tes I stars show Sr/Ba$\lesssim 1$. \citet{aoki05} 
found that some of the halo stars with [Fe/H]$<-3.0$ show Sr/Ba$>1$, 
while the excess of Sr is not seen for stars with higher 
Ba abundances. They interpret the result as evidence that 
the two astrophysical 
sites are responsible for producing these neutron-capture elements, 
one of which produces both Sr and Ba and the other 
produces more Sr than Ba. Since the signature of 
enhanced Sr relative to Ba is not evident in our 
Bo\"{o}tes I sample, the latter 
astrophysical site may be lacking or less frequent 
in this galaxy. We discuss the origin of Sr and Ba in Bootes I 
in Section \ref{sec:yields}.

To summarize, the present sample of Bo\"{o}tes I stars 
show a hint of difference in abundances of some of the  
$\alpha$-elements and neutron-capture elements from the bulk of 
MW halo stars. 
This result suggests that the chemical abundances of 
Bo\"{o}tes I stars alone are not enough to 
fully explain the observed chemical abundances for field halo stars. 

\citet{corlies13} have investigated conditions that yield 
a difference in chemical abundances 
between surviving satellite galaxies and progenitors of the MW 
halo using an N-body simulation. They suggest that the progenitors 
of the stellar halo usually experienced more cross pollution by 
neighboring subhalos, while the surviving satellites tend to have 
evolved chemically in isolation. Whether the degree of the 
cross pollution explains the observed abundance difference may 
have interesting implications for the hierarchical formation of the 
MW halo.

Since the present result is based on a small sample, 
a larger sample is clearly desirable to extract a  
conclusion for the range in elemental abundances spanned 
by the whole stellar population in Bo\"{o}tes I. 
Moreover, abundances of field halo stars are often determined only 
for relatively nearby objects (within a several kpc), 
while the estimated distance to Bo\"{o}tes I is 64$\pm 3$ 
kpc \citep{okamoto12}. A systematic abundance survey 
for more distant halo stars would be needed to examine 
possible connection between dSphs similar to 
Bo\"{o}tes I and the MW field halo.

\subsection{Comparison with other dSphs}
\label{sec:otherdsph}

\begin{figure}
   \centering
   \includegraphics[width=8.5cm]{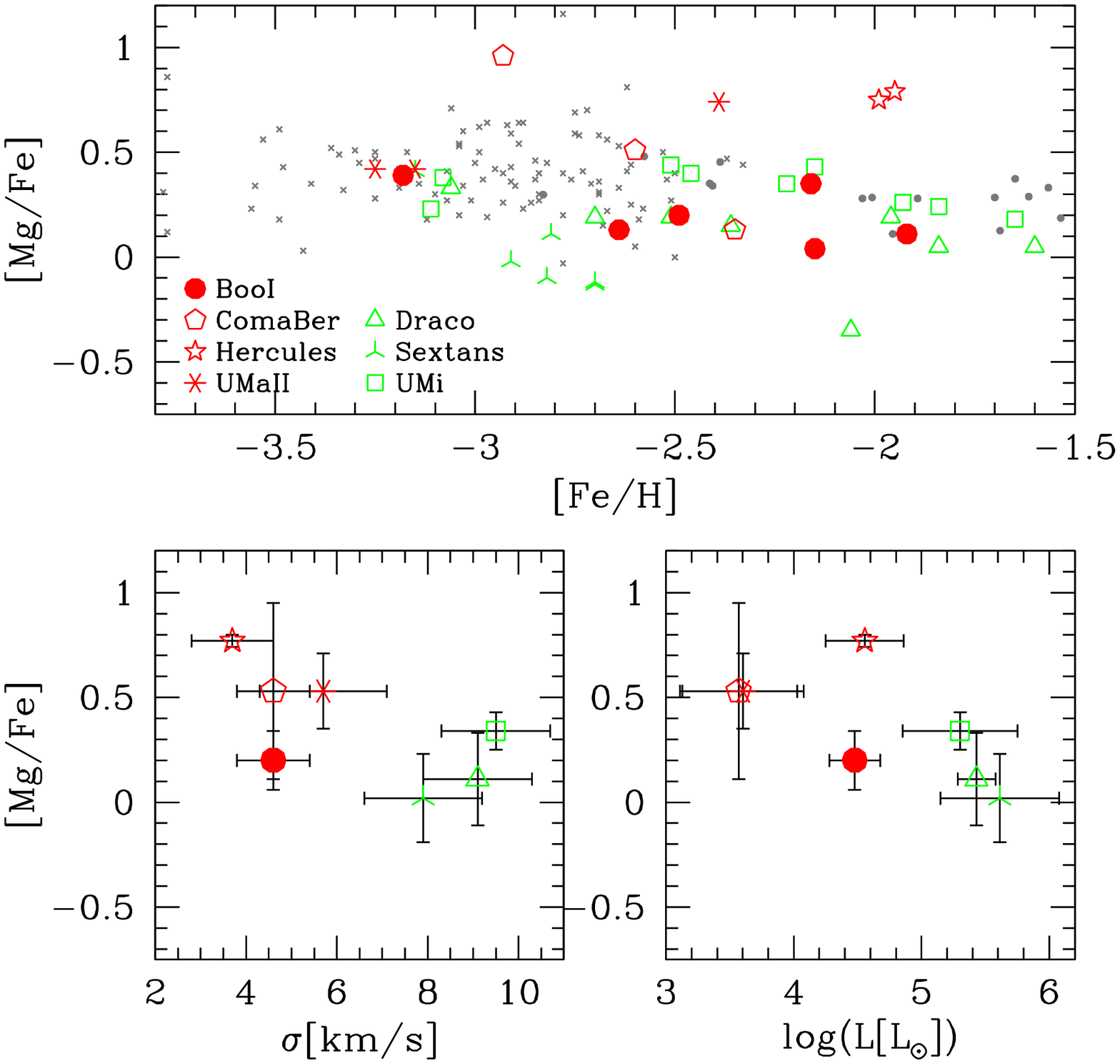}
   \caption{Top panel: Abundance comparisons with classical (Draco: triangles, Sextans: three pointed asterisk, Ursa Minor: square)  and ultra-faint (Coma Berenices: pentagon, Hercules: star, Ursa Major II: asterisk) dwarf galaxies 
from the literature \citep{aoki09,cohen09,cohen10,frebel10,koch08}. For comparison, the abundance data for MW halo giants extracted from SAGA database \citep{suda08} ($R>30000$ with no flags of ``C-rich" or ``CEMP'') and from \citet{ishigaki12}  are plotted with crosses and dots, respectively. Bottom panels: Mean [Mg/Fe] ratios for 
each galaxy plotted against velocity dispersion ($\sigma$, left) and luminosity ($\log L(L_{\odot})$, right) for that galaxy. The $\sigma$ and $L$ data are 
taken from \citet{walker09}.}
   \label{fig:mgfe}
   \end{figure}

Figure \ref{fig:mgfe} compares the [Mg/Fe] ratios of the Bo\"{o}tes I stars 
in this work and those of other MW dSphs that include 
very metal-poor ([Fe/H]$<-2$) stars from the literature 
\citep{aoki09,cohen09,cohen10,frebel10,koch08}. Relatively 
luminous ($L_{V, \odot}\gtrsim 10^{5} L_{\odot}$), so-called classical 
dSphs (Draco, Sextans, Ursa Minor) are shown with green symbols, 
while less luminous dSphs (UFDs; Bo\"{o}tes I, Coma Berenices, Hercules, 
Ursa Major II) are shown in red symbols. 
 According to these available data, 
the more luminous dSphs appear to show relatively low [Mg/Fe] ratios 
than field halo stars at [Fe/H]$\sim -2$ 
and/or a decreasing [Mg/Fe] trend 
with increasing [Fe/H].  On the other hand, the fainter dSphs such as 
Coma Berenices or Ursa Major II appear to show comparable 
or higher abundance ratios with large scatter. 
The Bo\"{o}tes I stars generally show 
the [Mg/Fe] ratios more similar to 
some of the classical dSphs, such as Ursa Minor rather than the 
fainter UFDs, such 
as Coma Berenices or Ursa Major II, although there seems 
to be large variation in the abundance ratios among 
the classical dSphs.  

\begin{figure}
   \centering
   \includegraphics[width=8.5cm]{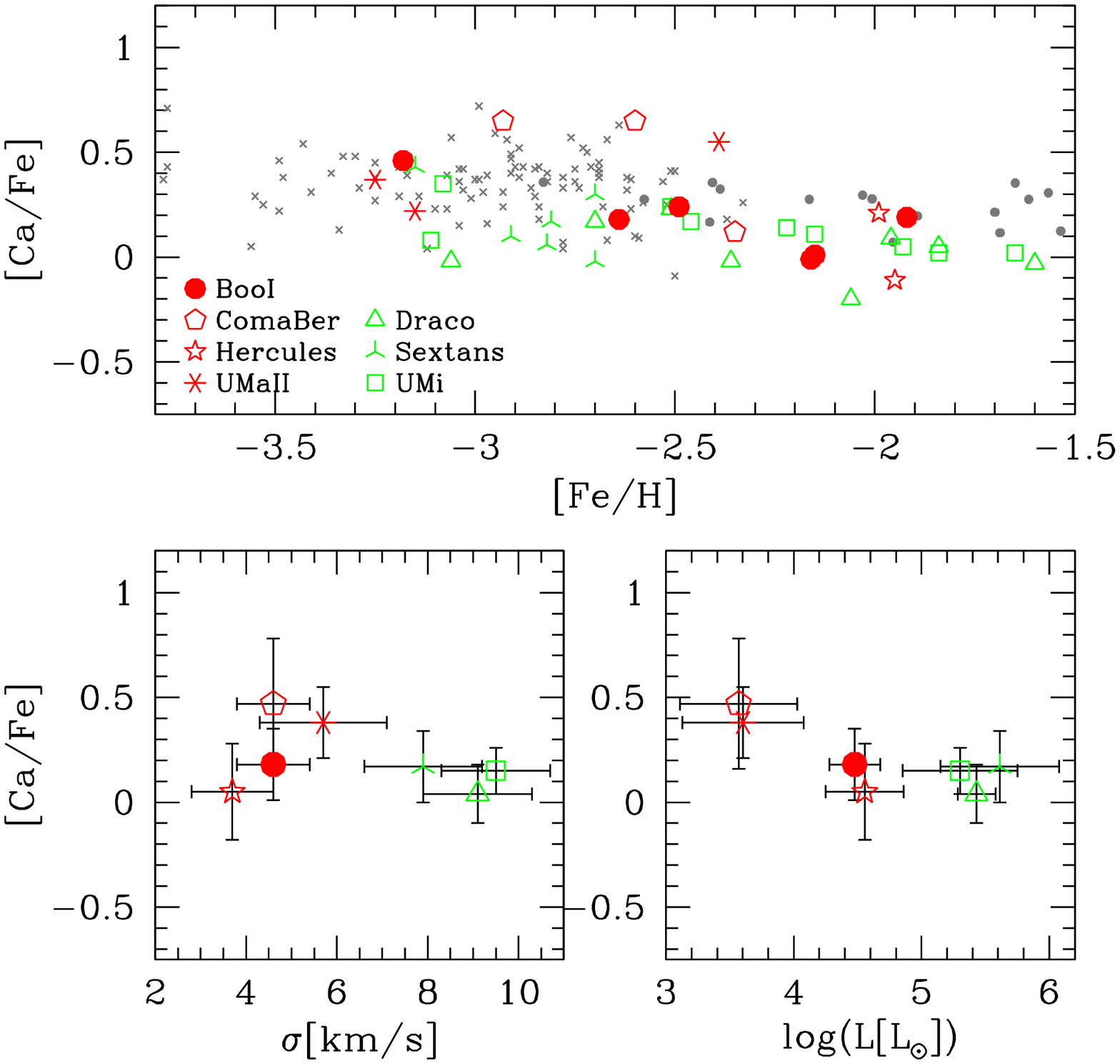}
   \caption{Same as Figure \ref{fig:mgfe} but for [Ca/Fe].}
   \label{fig:cafe}
   \end{figure}

To examine whether the abundance ratios 
can be related to the global properties of dSphs 
(e.g., dynamical masses or luminosity),  
the lower panels in Figure \ref{fig:mgfe} plot average [Mg/Fe] ratios
of these dSphs against their central line-of-sight velocity 
dispersions ($\sigma_{\rm los}$; 
left) and logarithms of total luminosity ($L$; right). 
The data for $\sigma$ and $L$ are taken from 
\citet{walker09} except for the $\sigma$ of Bo\"{o}tes I, for which 
an updated value from \citet{koposov11} ($4.6^{+0.8}_{-0.6}$ km s$^{-1}$, 
assuming a single component velocity distribution) is available. 
These figures suggest that the dSphs with larger
 velocity dispersions or higher luminosity tend 
to show lower [Mg/Fe] ratios. 
Again, Bo\"{o}tes I, which has $\sigma$ as low as that of 
Coma Berenices and $L$ midway between these two classes of dSphs, 
show the mean [Mg/Fe] ratio to be more similar to the classical 
dSphs, such as Ursa Minor or Draco. 
A similar trend can be seen for [Ca/Fe] ratios
as shown in Figure \ref{fig:cafe}.
The exceptionally 
high [Mg/Fe] ratios are reported for the two 
Hercules stars, which may suggest an anomalous star formation 
history for this galaxy \citep[see also][]{Francois12}. 
According to the available data, differences 
in other elemental abundances lighter than zinc, 
such as [Na/Fe] ratios, are not very clear among all dSphs including 
Bo\"{o}tes I for
[Fe/H]$<-2$. 

The MW dSphs 
have been reported to follow a correlation between 
mean [Fe/H] and $L$ \citep{kirby11a} extending 
for a wide range of luminosities ($3.5<\log (L/L_{\odot})<7.5$). 
Although precision and sample size may not be sufficient 
to examine the [Mg,Ca/Fe]-$\sigma$, -$L$ relation, if any, from 
currently available data,  
the locations of Bo\"{o}tes I in these diagrams may have implications 
for the chemical evolution of this galaxy, which we
discuss in Section \ref{sec:chemevol}

\begin{figure}
   \centering
   \includegraphics[width=8.5cm]{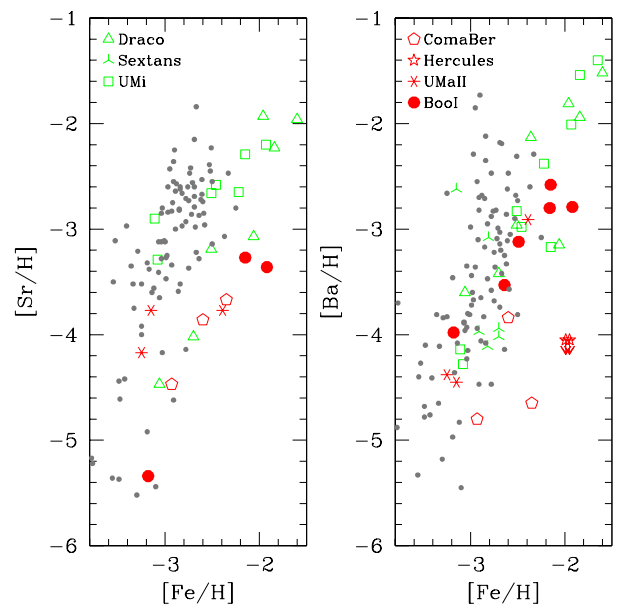}
   \caption{Comparison of the [Sr/H] and [Ba/H] ratios in Bo\"{o}tes I stars 
obtained in this work and those in very metal-poor stars in other dSphs. 
Symbols are the same as in Figure \ref{fig:mgfe}.}
   \label{fig:sr_ba_feh}
\end{figure}

Figure \ref{fig:sr_ba_feh} shows [Sr/H] and [Ba/H] ratios 
plotted against [Fe/H] for Bo\"{o}tes I and other dSphs. 
Classical dSphs generally show increasing [Sr/H] and 
[Ba/H] ratios with increasing [Fe/H], similar to the 
trend seen for field halo stars.
On the other hand, the abundances for 
fainter dSphs tend to distribute at lower bounds than for the 
field halo stars. 
 Lower abundances of neutron-capture elements for very-metal-poor 
stars in dSphs have already been noticed in  
previous studies \citep[e.g.,][]{fulbright04,koch08,frebel10,honda11,koch13}. 
Our sample of Bo\"{o}tes I  
stars show [Ba/H] ratios that are more similar to the classical dSphs, 
while [Sr/H] ratios are lower, similar to those observed in 
fainter dSphs.  

Whether the neutron-capture elements are more deficient in dSphs 
than in the MW field halo stars needs to be investigated with a 
larger sample of high-quality spectra for distant objects, which 
may be challenging with 
the current instrumentation. 
Nevertheless, the detection of Sr and Ba in Bo\"{o}tes I suggests that at 
least one event that produces neutron-capture elements 
must have occurred at the earliest epoch of star formation in 
this galaxy \citep[cf.,][]{roederer13}.

\subsection{Chemical evolution of Bo\"{o}tes I}
\label{sec:chemevol}

In Section \ref{sec:compmw}, we reported a possible offset in 
the abundances of $\alpha$ and neutron capture elements between 
the Bo\"{o}tes I stars and the field halo stars 
in the metallicity range $-3<$[Fe/H]$<-2$. 
These differences, if confirmed with larger samples, 
may suggest the differences in chemical evolution history 
between the Bo\"{o}tes I and dominant progenitors of the MW stellar 
halo. As discussed in Section \ref{sec:otherdsph}, the [Mg/Fe] and [Ca/Fe] ratios in the
observed Bo\"{o}tes I stars are generally similar to the values 
reported for brighter (classical) dSphs in the metallicity range 
$-3<$[Fe/H]$<-2$, while their velocity dispersion 
and luminosities are more similar to those of UFDs 
 (Figures \ref{fig:mgfe} and \ref{fig:cafe}). 

In this section, we discuss possible implications 
of the results of these comparisons on the 
chemical evolution of Bo\"{o}tes I
in terms of 1) contribution from Type Ia SNe, 2) galactic winds, 
3) tidal interaction with the MW,  
4) stochastic effects, and 5) individual supernova yields.

\subsubsection{Contribution from Type Ia SNe}

In chemical evolution models, trends in abundances 
of $\alpha$-elements with [Fe/H] are diagnostics of 
the Type Ia SNe contribution relative to Type II SNe to 
chemical enrichments in galaxies \citep[e.g.,][]{matteucci86}.
For many of known dSphs, decreasing trends of [$\alpha$/Fe] with 
increasing [Fe/H] are reported \citep{tolstoy09,
kirby11b,vargas13,kirby13}, which 
could be indicative of chemical enrichment proceeded 
partly through Type Ia SNe. 

Whether Bo\"{o}tes I has experienced a significant 
contribution from Type Ia SNe to its chemical enrichments 
is an important question for 
constraining the star formation history in this galaxy. 
Photometric observations suggest that Bo\"{o}tes I is dominated by
an old stellar population with little age spread \citep{okamoto12}, 
which suggests that the duration of star formation 
was shorter than the uncertainty of the photometric 
age estimates (a few Gyr).   
\citet{gilmore13} suggest a contribution of Type Ia SNe in 
order to interpret the apparent decreasing [$\alpha$/Fe] trend with [Fe/H]. 
On the other hand, the low-resolution spectroscopic 
study of \citet{lai11} does not find a signature of decreasing 
[$\alpha$/Fe] ratios with increasing [Fe/H], which does not support 
the Type Ia contribution.

The observed [Mg/Fe] and [Ca/Fe] ratios for the Bo\"{o}tes I stars 
in this work appear to show a modest decreasing trend with [Fe/H]. 
The Pearson test for linear correlation indicates that the 
probability for a null correlation is 30 \% for [Mg/Fe]-[Fe/H] 
and 7 \% for [Ca/Fe]-[Fe/H], thus only marginally supporting
the correlation. 
Although a firm conclusion about the trend could not be 
obtained for the present sample, 
the hint of the decreasing trend suggests that 
the star formation in Bo\"{o}tes I lasted at least until Type Ia SNe 
began to contribute to the chemical enrichment. 

The decreasing [Ca/Fe] trend with [Fe/H] has also been reported for 
a sample of red giant stars in Hercules dSph, which spans the [Fe/H] range 
similar to the present Bo\"{o}tes I sample. 
This may imply that the contribution of Type Ia SNe to the 
chemical enrichment 
is common among galaxies as faint as Hercules or Bo\"{o}tes I, if 
the Type Ia SNe is indeed responsible for lowering [Ca/Fe]. 
Analyses of the delay-time distribution of Type Ia 
SNe suggest that a certain fraction of Type Ia SNe 
occur within $<1$ Gyr of star formation \citep{maoz10}, 
which may support their contribution to the 
chemical enrichments in Bo\"{o}tes I. 

An alternative interpretation of the lower [Mg, Ca/Fe] ratios in 
Bo\"{o}tes I is a different initial mass function (IMF)
 in this galaxy in comparison to the dominant progenitors 
of the MW halo.  Since [Mg/Fe] yields in Type II SNe 
could depend on masses of the progenitor stars \citep{kobayashi06}, 
a difference in 
IMF may lead to systematic difference in these abundance ratios.

\subsubsection{Galactic wind} 

Galactic winds driven by SNe are thought to significantly affect 
subsequent chemical evolutions in dSphs \citep{arimoto87}. Chemical 
evolution models of \citet{lanfranchi04} particularly suggest that  
the strength and time of occurance of the galactic winds 
play an important role in determining the [$\alpha$/Fe] ratios in 
the dSphs.     
  
The winds are expected to occur when part of the energy injected 
by SNe heat up interstellar gas so that thermal 
energy of gas exceed the binding energy of the galaxy \citep{arimoto87}. 
In this scenario, the strength of the wind may depend on 
both the initial gas content and gravitational potential of the 
galaxy. 

As can be seen in the lower left-hand panels 
of Figures \ref{fig:mgfe} and \ref{fig:cafe}, the dSphs 
with smaller velocity dispersion, and thus lower dynamical 
masses, such as Coma Berenices or Ursa Major II, show higher [Mg/Fe] 
and [Ca/Fe] ratios than the dSphs with larger velocity dispersions. 
The likely correlation between the velocity dispersion 
and the abundance ratios suggests that the chemical abundance 
depends on the gravitational potential of 
galaxies, if the velocity dispersion traces the 
dynamical mass of these systems. 

Using the measured velocity dispersion 
profile, \citet{walker09} estimate the dynamical mass of 
Bo\"{o}tes I within the half-light radius of this galaxy 
($r_{\rm half}=242\pm 21$ pc) to be $\sim 10^6 M_{\odot}$. 
If we assume that the mass has not 
changed significantly over the age of this galaxy 
and if duration of star formation is indeed longer than the 
timescale of Type Ia SNe, as suggested in the previous section, 
then the above scenario implies that a dark matter halo with 
mass as low as that of Bo\"{o}tes I could 
host enough initial gas to retain a certain duration of 
star formation, or that galactic winds were not strong enough to truncate 
all subsequent star formation in such a halo.
Whether this scenario could be possible depends on 
the possible tidal effect on the initial gas and dark matter 
content of this galaxy as discussed below.   

\subsubsection{Tidal Interaction with the MW}
\label{sec:tidal}

We discussed in Section \ref{sec:otherdsph} a possible 
connection of chemical abundance ratios with global properties, such as 
$\sigma_{\rm los}$ and $L$. The correlation is not currently clear, and 
Bo\"{o}tes I in particular appears to deviate from a possible relation. 
Thus, we may ask what makes Bo\"{o}tes I smaller $\sigma_{\rm los}$ 
and lower $L$, while experiencing chemical evolution that is 
more like the
brighter classical dSphs. 

One possible mechanism that may affect $\sigma_{\rm los}$ or $L$ is tidal 
interaction with the MW potential. 
Tidal interaction of Bo\"{o}tes I with 
the MW may have stripped the original 
content of baryons and dark matter in this galaxy. 
N-body simulations for dynamical evolution 
of Bo\"{o}tes I suggest that this galaxy has lost a significant 
amount of luminous and dark matter contents by the tidal forces 
as it orbits around the MW potential \citep{fellhauer08}.  
It is possible that Bo\"{o}tes I was initially as luminous as 
the classical dSphs, while it has lost a significant amount of 
materials.

Since the full orbital motion of Bo\"{o}tes I is not known, 
it is difficult to 
estimate how strong the tidal effects were in this galaxy.
The tidal radius of this galaxy at its present location ($62$ kpc 
from the Galactic center) is estimated to be $1.2$ kpc \citep{fellhauer08}, 
which is larger than the luminous component in this galaxy. Therefore, 
it is unlikely that the tidal effect is currently significant in stripping 
material from this galaxy. 
On the other hand, 
\citet{collins13} report that the estimated 
maximum circular velocity ($V_{\rm max}$) for Bo\"{o}tes I is relatively low compared to
other local group dSphs with wide ranges of luminosities, which could 
be explained by tidal stripping of dark matter.
Therefore, one explanation of the observed 
similarity may be that Bo\"{o}tes I once experienced similar 
chemical evolution to, say, Ursa Minor, then within a few Gyr, 
it followed an orbit that brought it closer to the Galactic center, 
experiencing strong tidal force that were able to strip luminous and 
dark matter contents in this galaxy. All subsequent star formation 
stopped, which results in the current 
low luminosity and very low metallicity ([Fe/H]$<-1.5$) 
in this galaxy \citep{norris10b}.
The information about the orbital motion of Bo\"{o}tes I would be 
necessary for examining such a scenario to interpret the 
observed chemical abundances in this galaxy.

\subsubsection{Stochastic effect}

In very low-metallicity systems like Bo\"{o}tes I, 
it is possible that  
only a small number of SNe of a prior generation of stars 
have contributed to the chemical enrichment. 
For example, \citet{frebel12} suggest that first galaxies 
that formed exclusively from ejecta of the first generation of stars 
(Pop III stars)
are expected to host relatively 
few progenitor SNe, which could be around ten, compared 
to systems with higher metallicity. 
In this case, the effects of stochastic sampling of 
an IMF may become important. 

The present sample of Bo\"{o}tes I stars are rather 
homogeneous in abundance ratios for elements lighter than zinc, which 
does not strongly support the existence of any stochastic effect. 
Nevertheless, the observed homogeneous [X/Fe] ratios 
do not completely rule out the 
stochastic enrichment scenario, since little scatter in 
[X/Fe] is expected, while producing large scatter in [Fe/H] 
in such scenarios as proposed by \citet{frebel12}. 
This scenario thus implies that Bo\"{o}tes I could be a candidate for
the first galaxies formed in the Universe. 

\citet{lee13} have recently investigated the effect of the 
stochastic IMF sampling as a 
possible interpretation 
of the abundance difference and similarity
between the UFDs and MW halo stars. 
In their model, a system enriched by a smaller number of 
a prior generation of stars is expected to show a 
lower mean value of [X/Fe] with 
larger spread, when the supernova yield of an element X has a positive 
mass dependence (i.e., higher ejected mass for 
a higher supernova progenitor mass).  
Therefore, if a UFD has been enriched by fewer
 SNe ($\sim 1-7$ on average) than the progenitors 
of the MW halo ($>7$) and if a Sr yield, which is currently 
uncertain, has a strong positive mass dependence, 
we would expect lower [Sr/Fe] for the UFDs than the MW halo progenitor. 
They conclude that the stochastic sampling of 
IMF for UFDs can simultaneously explain the observed difference in 
[Sr, Ba/Fe] ratios and the similarity 
in [Ti/Fe] ratios between the UFDs and MW halo stars.

The result of the present work, a modest difference in [Mg/Fe] 
and a large difference in [Sr/Fe] from field halo stars 
appear to both be reproduced 
with the above model if the Sr yield 
 depends more strongly on progenitor masses than the Mg yields do. 
Thus, according to this model, the difference in number of 
prior-generation stars can explain the observed abundance difference 
between the Bo\"{o}tes I and progenitors of the MW halo.

\subsubsection{Individual yields}
\label{sec:yields}

We finally discuss whether the abundances of the present sample stars 
are consistent with yields of a single SN of Pop III stars. 
Pop III stars may have been very massive. Very 
massive stars with M$>140 M_{\odot}$ are predicted to be 
exploded as pair-instability SNe whose yields are expected to show a
peculiar abundance pattern, including very large Ca abundances. 
As can be seen, none of our sample stars show this peculiar 
pattern and seem to be more consistent with yields 
of core-collapse SNe \citep[e.g.,][]{tominaga07}. 

An intriguing issue is whether nucleosynthetic 
yields of SNe are compatible with 
the observed abundances of Sr and Ba in Bo\"{o}tes I. 
Core-collapse SNe of massive ($>10M_{\odot}$) stars have been
considered as one of the promising production sites of heavy 
neutron-capture elements such 
as Ba \citep[e.g.,][]{thielemann11}.
 However, more recent studies report difficulty in reproducing 
the required physical condition for the r-process to occur 
in the environment 
of the core-collapse SNe. 

\citet{qian08} used the phenomenological 
model to study nucleosynthetic yields of neutron-capture 
elements in neutrino-driven winds, which are associated with SNe. They 
pointed out that the low [Sr/Fe] ratios are not 
reproduced with the predicted yields of Type II SNe with 
progenitor masses of neither $8-10 M_{\odot}$ (H source) nor
 $12-25 M_{\odot}$ (L sources). Instead, they suggest that 
another source that produces a large amount of Fe but 
little Sr is required to explain the abundance patterns of stars with 
[Sr/Fe]$<<-0.32$. They speculate that an energetic SN, 
a so-called hypernova, could be a possible astrophysical site that yields 
the low [Sr/Fe] ratios.
In this scenario, the observed 
Bo\"{o}tes I stars may have formed out of interstellar medium 
predominantly enriched by a single supernovae with 
large explosion energy. 

A merging of neutron stars in a close binary system is proposed 
as an alternative site of the r-process. 
Recent calculations of nucleosynthetic yields 
for the neutron-star merger successfully explain the abundance patterns 
of known r-process-rich stars in the Galactic field halo 
\citep[e.g.,][]{wanajo12,korobkin12}. 
It is interesting for future studies to investigate whether the 
predicted nucleosynthetic yields of neutron star mergers 
can explain the observed 
Sr/Fe and/or Ba/Fe abundances in  Bo\"{o}tes I. 

The supernova yields depend on various unknown parameters, such as
mass cut or explosion energies, and thus properties of the 
progenitor SN are not fully constraind at this moment. 
Nucleosynthetic yields of light and heavy neutron capture elements for
different SN progenitor masses at extremely low metallicities
are required to examine the reason for 
the Sr deficiency in Bo\"{o}tes I.

%

\section{Conclusions \label{sec:conclude}}
The chemical abundances of C, Na, $\alpha$, iron-peak, and 
        neutron capture elements in six giant stars in Bo\"{o}tes I
        ultra-faint dwarf galaxy were estimated based on the 
        high-resolution spectra obtained with Subaru/HDS. Our 
main results are summarized as follows: 

   \begin{enumerate}
      \item 
        The very low metallicity of Boo--094 ([Fe/H]$=-3.4$) previously 
        reported in \citet{feltzing09} and \citet{gilmore13} is confirmed.
        For the six sample stars, extremely carbon-rich objects are 
        not found.
      \item For $\alpha$ and iron-peak elements, 
        the scatter in the abundance ratios is small 
        over the metallicity range of $-2.7<$[Fe/H]$<-1.8$.   
      \item The [Mg/Fe] and [Ca/Fe] ratios for these stars are 
            slightly lower than the field halo stars on average 
            in the metallicity range of $-3.0<$[Fe/H]$<-2.0$. 
            On the other hand, the abundance ratios 
            for Na and Sc in the Bo\"{o}tes I 
           stars are generally similar to those of the field halo 
           stars. 
         \item The [Sr/Fe] abundances of the three Bo\"{o}tes I stars 
          are lower than the field halo values at similar 
           metallicities.  These Bo\"{o}tes I stars 
           deviate from the correlation 
           between $\log\epsilon$(Ba) and $\log\epsilon$(Sr) seen in 
           field MW halo stars and generally show lower Sr/Ba ratios. 
         
   \end{enumerate}

Highly inhomogeneous chemical abundances that are reported in fainter 
dSphs are not seen in our Bo\"{o}tes I sample. Instead, our  
results suggest that the star formation in 
Bo\"{o}tes I had lasted until Type Ia SNe started 
to contribute to the chemical enrichment, which has also been 
reported in other more luminous dSphs.
Larger samples of chemical abundances and the kinematics 
of individual stars, as well as a determination of the 
orbital motion around the MW,   
are desirable to conclude whether Bo\"{o}tes I is once more 
similar to brighter dSphs or is a surviving 
example of first galaxies formed out of ejecta from 
Pop III stars.

\begin{acknowledgements}
We thank the referee for constructive comments 
and suggestions that have improved our paper. 
We are grateful to S. Feltzing for providing us with their 
Mg linelist. M.N.I. thanks K. Nomoto, L. Vargas, and A. Frebel for 
helpful discussions on the abundance analysis 
and chemical evolution of 
ultra-faint dwarf galaxies.  
M.N.I. acknowledges financial support from Grant-in-Aid for 
JSPS (Japan Society for the Promotion of Science) fellow. 
W.A. was supported by the JSPS Grants-in-Aid for 
Scientific Research (23224004).
\end{acknowledgements}

\bibliography{boo}
\bibliographystyle{aa}

\onecolumn
\begin{table}
\caption{\label{tab:ew}Equivalent widths and abundances. Table 4 is published in its entirety in the electronic edition of the Astronomy \& Astrophysics. A portion is shown here for guidance regarding its form and context.}
\begin{center}
\begin{tabular}{lccccccc}
\hline\hline
Object & Elem. & Ion & Wavelength & $\chi$ & $\log gf$ & EW & $\log\epsilon A$ \\ 
 &   &   & ({\AA})& (eV) & & (m{\AA}) & (dex) \\ 
\hline
   Boo--009 & 11 & 1 &  5895.92 &   0.00 &   $-$0.19 &   150.0 &    3.61 \\ 
   Boo--009 & 12 & 1 &  4702.99 &   4.35 &   $-$0.44 &    50.1 &    4.94 \\ 
   Boo--009 & 12 & 1 &  5528.40 &   4.35 &   $-$0.50 &    69.7 &    5.24 \\ 
\hline
\end{tabular}
\end{center}
\end{table}

\begin{longtable}{lccccccccccc}
\caption{\label{tab:abund}Elemental abundances and uncertainties}\\
\hline\hline
Name & Elem. & $\log\epsilon A_{\odot}$ & $\log\epsilon A$& [X/H] & [X/Fe] & N & $\sigma_{\rm line}$ & $\Delta^{+100 {\rm K}}_{-100 {\rm K}}$ & $\Delta^{+0.3 {\rm dex}}_{-0.3 {\rm dex}}$&  $\Delta^{+0.3 {\rm km s}^{-1}}_{-0.3 {\rm km s}^{-1}}$ & $\sigma_{\rm tot}$ \\
\hline
\endfirsthead
\caption{continued.}\\
\hline\hline
Name & Elem. & $\log\epsilon A_{\odot}$ & $\log\epsilon A$& [X/H] & [X/Fe] & N & $\sigma_{\rm line}$ & $\Delta^{+100 {\rm K}}_{-100 {\rm K}}$ & $\Delta^{+0.3 {\rm dex}}_{-0.3 {\rm dex}}$ &  $\Delta^{+0.3 {\rm km s}^{-1}}_{-0.3 {\rm km s}^{-1}}$  & $\sigma_{\rm tot}$ \\
\hline
\endhead
\hline
\endfoot
     Boo--009 &   FeI &    7.50 &    4.86 &  $-$2.64 &  $-$2.64 & 38 &    0.04 & $^{   0.12}_{  -0.12}$ & $^{  -0.02}_{   0.03}$ & $^{  -0.04}_{   0.06}$ &   0.14 \\ 
            &  FeII &    7.50 &    4.59 &  $-$2.91 &  $-$2.91 &  2 &    0.01 & $^{   0.01}_{   0.00}$ & $^{   0.10}_{  -0.09}$ & $^{  -0.01}_{   0.02}$ &   0.10 \\ 
            & C(CH) &    8.43 &$<$   5.50 &$< -$2.93 &$< -$0.29 &  1 & ... & ... & ... & ... & ... \\ 
            &   NaI &    6.24 &    3.61 &  $-$2.63 &    0.01 &  1 &    0.25 & $^{   0.01}_{  -0.01}$ & $^{  -0.03}_{   0.02}$ & $^{  -0.10}_{   0.10}$ &   0.27 \\ 
            &   MgI &    7.60 &    5.09 &  $-$2.51 &    0.13 &  2 &    0.15 & $^{  -0.06}_{   0.06}$ & $^{   0.00}_{  -0.00}$ & $^{   0.01}_{  -0.02}$ &   0.16 \\ 
            &   CaI &    6.34 &    3.88 &  $-$2.46 &    0.18 &  2 &    0.04 & $^{  -0.03}_{   0.04}$ & $^{   0.00}_{  -0.00}$ & $^{   0.00}_{  -0.01}$ &   0.05 \\ 
            &  ScII &    3.15 &    0.39 &  $-$2.76 &    0.15 &  2 &    0.09 & $^{   0.04}_{  -0.04}$ & $^{  -0.00}_{   0.01}$ & $^{   0.00}_{  -0.00}$ &   0.09 \\ 
            &   TiI &    4.95 &    2.56 &  $-$2.39 &    0.24 &  2 &    0.07 & $^{   0.01}_{  -0.01}$ & $^{  -0.00}_{   0.00}$ & $^{   0.02}_{  -0.02}$ &   0.07 \\ 
            &  TiII &    4.95 &    2.31 &  $-$2.64 &    0.27 &  2 &    0.11 & $^{   0.04}_{  -0.04}$ & $^{  -0.01}_{   0.01}$ & $^{  -0.02}_{   0.03}$ &   0.12 \\ 
            &   CrI &    5.64 &    2.25 &  $-$3.39 &  $-$0.75 &  1 &    0.25 & ...& $^{  -0.00}_{   0.00}$ & ...&   0.25 \\ 
            &   NiI &    6.22 &    3.60 &  $-$2.62 &    0.02 &  2 &    0.07 & $^{   0.00}_{  -0.00}$ & $^{   0.00}_{  -0.01}$ & $^{   0.01}_{  -0.01}$ &   0.07 \\ 
            &   ZnI &    4.56 &$<$   2.20 &$< -$2.36 &$<$   0.28 &  1 & ... & ... & ... & ... & ... \\ 
            &  SrII &    2.87 & ... & ... & ... &  0 & ... & ... & ... & ... & ... \\ 
            &  BaII &    2.18 &  $-$1.61 &  $-$3.79 &  $-$0.89 &  1 &    0.25 & $^{   0.08}_{  -0.08}$ & $^{  -0.01}_{   0.01}$ & $^{  -0.02}_{   0.02}$ &   0.27 \\ 
\hline
     Boo--094 &   FeI &    7.50 &    4.32 &  $-$3.18 &  $-$3.18 & 22 &    0.04 & $^{   0.12}_{  -0.12}$ & $^{  -0.04}_{   0.04}$ & $^{  -0.02}_{   0.03}$ &   0.14 \\ 
            &  FeII &    7.50 &    4.02 &  $-$3.48 &  $-$3.48 &  2 &    0.16 & ...& $^{   0.09}_{  -0.08}$ & $^{  -0.03}_{   0.03}$ &   0.19 \\ 
            & C(CH) &    8.43 &$<$   5.50 &$< -$2.93 &$<$   0.25 &  1 & ... & ... & ... & ... & ... \\ 
            &   NaI &    6.24 &    2.73 &  $-$3.51 &  $-$0.32 &  1 &    0.19 & $^{  -0.01}_{   0.00}$ & $^{  -0.01}_{   0.01}$ & $^{  -0.07}_{   0.08}$ &   0.20 \\ 
            &   MgI &    7.60 &    4.81 &  $-$2.79 &    0.39 &  3 &    0.05 & $^{  -0.03}_{   0.03}$ & $^{   0.00}_{  -0.00}$ & $^{   0.01}_{  -0.01}$ &   0.06 \\ 
            &   CaI &    6.34 &    3.61 &  $-$2.73 &    0.46 &  2 &    0.01 & $^{  -0.04}_{   0.05}$ & $^{   0.01}_{  -0.01}$ & $^{   0.01}_{  -0.01}$ &   0.05 \\ 
            &  ScII &    3.15 & ... & ... & ... &  0 & ... & ... & ... & ... & ... \\ 
            &   TiI &    4.95 &    1.79 &  $-$3.16 &    0.02 &  1 &    0.19 & $^{   0.01}_{  -0.01}$ & $^{   0.00}_{  -0.00}$ & $^{   0.01}_{  -0.02}$ &   0.19 \\ 
            &  TiII &    4.95 &    2.02 &  $-$2.93 &    0.55 &  1 &    0.19 & $^{   0.04}_{  -0.04}$ & $^{  -0.00}_{  -0.00}$ & $^{   0.02}_{  -0.03}$ &   0.19 \\ 
            &   CrI &    5.64 &    1.94 &  $-$3.70 &  $-$0.52 &  1 &    0.19 & $^{   0.00}_{  -0.00}$ & $^{  -0.00}_{   0.00}$ & $^{  -0.00}_{   0.00}$ &   0.19 \\ 
            &   NiI &    6.22 &    2.96 &  $-$3.26 &  $-$0.08 &  1 &    0.19 & $^{  -0.00}_{   0.00}$ & $^{   0.00}_{  -0.00}$ & $^{   0.00}_{  -0.00}$ &   0.19 \\ 
            &   ZnI &    4.56 &$<$   2.30 &$< -$2.26 &$<$   0.92 &  1 & ... & ... & ... & ... & ... \\ 
            &  SrII &    2.87 &  $-$2.47 &  $-$5.34 &  $-$2.16 &  1 &    0.70 & ...& ...& ...&   0.70 \\ 
            &  BaII &    2.18 &  $-$2.10 &  $-$4.28 &  $-$0.80 &  3 &    0.10 & $^{   0.07}_{  -0.08}$ & $^{  -0.01}_{   0.01}$ & $^{   0.01}_{  -0.01}$ &   0.12 \\ 
\hline
     Boo--117 &   FeI &    7.50 &    5.35 &  $-$2.15 &  $-$2.15 & 32 &    0.04 & $^{   0.13}_{  -0.13}$ & $^{  -0.03}_{   0.03}$ & $^{  -0.10}_{   0.13}$ &   0.18 \\ 
            &  FeII &    7.50 &    5.11 &  $-$2.39 &  $-$2.39 &  3 &    0.14 & $^{   0.00}_{   0.01}$ & $^{   0.10}_{  -0.10}$ & $^{  -0.04}_{   0.05}$ &   0.18 \\ 
            & C(CH) &    8.43 &    5.49 &  $-$2.94 &  $-$0.79 &  1 &    0.30 & $^{  -0.13}_{   0.13}$ & $^{   0.03}_{  -0.03}$ & $^{   0.10}_{  -0.13}$ &   0.35 \\ 
            &   NaI &    6.24 &    3.84 &  $-$2.40 &  $-$0.25 &  2 &    0.05 & $^{   0.02}_{  -0.02}$ & $^{  -0.05}_{   0.04}$ & $^{  -0.05}_{   0.02}$ &   0.08 \\ 
            &   MgI &    7.60 &    5.50 &  $-$2.10 &    0.04 &  3 &    0.13 & $^{  -0.03}_{   0.03}$ & $^{  -0.00}_{   0.00}$ & $^{   0.03}_{  -0.05}$ &   0.14 \\ 
            &   CaI &    6.34 &    4.21 &  $-$2.13 &    0.01 &  6 &    0.11 & $^{  -0.04}_{   0.05}$ & $^{   0.01}_{  -0.01}$ & $^{   0.05}_{  -0.07}$ &   0.13 \\ 
            &  ScII &    3.15 &    0.85 &  $-$2.30 &    0.08 &  3 &    0.09 & $^{   0.04}_{  -0.04}$ & $^{  -0.00}_{   0.00}$ & $^{   0.01}_{  -0.01}$ &   0.10 \\ 
            &   TiI &    4.95 &    2.81 &  $-$2.14 &    0.01 &  3 &    0.13 & $^{   0.01}_{  -0.00}$ & $^{   0.00}_{  -0.00}$ & $^{   0.04}_{  -0.06}$ &   0.14 \\ 
            &  TiII &    4.95 &    3.00 &  $-$1.95 &    0.44 &  1 &    0.20 & $^{   0.04}_{  -0.04}$ & $^{  -0.01}_{   0.01}$ & $^{  -0.02}_{   0.03}$ &   0.20 \\ 
            &   CrI &    5.64 &    3.41 &  $-$2.23 &  $-$0.08 &  1 &    0.20 & $^{  -0.01}_{   0.02}$ & $^{   0.01}_{  -0.01}$ & $^{   0.08}_{  -0.10}$ &   0.22 \\ 
            &   NiI &    6.22 &    4.11 &  $-$2.11 &    0.04 &  1 &    0.20 & $^{   0.01}_{  -0.01}$ & $^{  -0.01}_{   0.01}$ & $^{  -0.06}_{   0.07}$ &   0.21 \\ 
            &   ZnI &    4.56 &$<$   2.50 &$< -$2.06 &$<$   0.09 &  1 & ... & ... & ... & ... & ... \\ 
            &  SrII &    2.87 &  $-$0.40 &  $-$3.27 &  $-$1.12 &  1 &    1.45 & ...& ...& ...&   1.45 \\ 
            &  BaII &    2.18 &  $-$0.64 &  $-$2.82 &  $-$0.43 &  2 &    0.05 & $^{   0.07}_{  -0.07}$ & $^{  -0.00}_{   0.00}$ & $^{  -0.03}_{   0.05}$ &   0.10 \\ 
\hline
     Boo--121 &   FeI &    7.50 &    5.01 &  $-$2.49 &  $-$2.49 & 34 &    0.03 & $^{   0.15}_{  -0.16}$ & $^{  -0.05}_{   0.05}$ & $^{  -0.08}_{   0.11}$ &   0.19 \\ 
            &  FeII &    7.50 &    4.89 &  $-$2.61 &  $-$2.61 &  4 &    0.07 & $^{  -0.00}_{   0.02}$ & $^{   0.09}_{  -0.08}$ & $^{  -0.06}_{   0.08}$ &   0.13 \\ 
            & C(CH) &    8.43 &$<$   5.70 &$< -$2.73 &$< -$0.24 &  1 & ... & ... & ... & ... & ... \\ 
            &   NaI &    6.24 &    3.50 &  $-$2.74 &  $-$0.25 &  2 &    0.05 & $^{   0.03}_{  -0.05}$ & $^{  -0.03}_{   0.02}$ & $^{  -0.08}_{   0.05}$ &   0.10 \\ 
            &   MgI &    7.60 &    5.31 &  $-$2.29 &    0.20 &  1 &    0.18 & $^{  -0.08}_{   0.09}$ & $^{   0.00}_{  -0.00}$ & $^{   0.02}_{  -0.04}$ &   0.20 \\ 
            &   CaI &    6.34 &    4.09 &  $-$2.25 &    0.24 &  4 &    0.06 & $^{  -0.05}_{   0.05}$ & $^{   0.00}_{  -0.01}$ & $^{   0.02}_{  -0.02}$ &   0.08 \\ 
            &  ScII &    3.15 &    0.41 &  $-$2.73 &  $-$0.13 &  3 &    0.11 & $^{   0.04}_{  -0.04}$ & $^{   0.00}_{  -0.01}$ & $^{   0.04}_{  -0.05}$ &   0.13 \\ 
            &   TiI &    4.95 &    2.59 &  $-$2.36 &    0.12 &  2 &    0.19 & $^{   0.01}_{  -0.02}$ & $^{  -0.00}_{   0.01}$ & $^{   0.03}_{  -0.03}$ &   0.19 \\ 
            &  TiII &    4.95 &    2.41 &  $-$2.54 &    0.07 &  1 &    0.18 & $^{   0.04}_{  -0.04}$ & $^{   0.00}_{  -0.00}$ & $^{   0.03}_{  -0.04}$ &   0.19 \\ 
            &   CrI &    5.64 & ... & ... & ... &  0 & ... & ... & ... & ... & ... \\ 
            &   NiI &    6.22 &    3.36 &  $-$2.86 &  $-$0.37 &  2 &    0.02 & $^{  -0.00}_{   0.01}$ & $^{   0.01}_{  -0.01}$ & $^{   0.03}_{  -0.04}$ &   0.04 \\ 
            &   ZnI &    4.56 &$<$   2.00 &$< -$2.56 &$< -$0.07 &  1 & ... & ... & ... & ... & ... \\ 
            &  SrII &    2.87 & ... & ... & ... &  0 & ... & ... & ... & ... & ... \\ 
            &  BaII &    2.18 &  $-$1.06 &  $-$3.24 &  $-$0.63 &  1 &    0.18 & $^{   0.07}_{  -0.08}$ & $^{   0.00}_{  -0.01}$ & $^{  -0.01}_{   0.02}$ &   0.20 \\ 
\hline
     Boo--127 &   FeI &    7.50 &    5.58 &  $-$1.92 &  $-$1.92 & 42 &    0.03 & $^{   0.15}_{  -0.15}$ & $^{  -0.03}_{   0.03}$ & $^{  -0.12}_{   0.14}$ &   0.21 \\ 
            &  FeII &    7.50 &    5.58 &  $-$1.92 &  $-$1.92 &  6 &    0.07 & $^{  -0.01}_{   0.02}$ & $^{   0.10}_{  -0.10}$ & $^{  -0.10}_{   0.13}$ &   0.17 \\ 
            & C(CH) &    8.43 &    5.74 &  $-$2.69 &  $-$0.77 &  1 &    0.30 & $^{  -0.15}_{   0.15}$ & $^{   0.03}_{  -0.03}$ & $^{   0.12}_{  -0.14}$ &   0.36 \\ 
            &   NaI &    6.24 &    4.14 &  $-$2.10 &  $-$0.18 &  2 &    0.14 & $^{   0.02}_{  -0.02}$ & $^{  -0.06}_{   0.06}$ & $^{  -0.02}_{  -0.02}$ &   0.15 \\ 
            &   MgI &    7.60 &    5.79 &  $-$1.81 &    0.11 &  2 &    0.01 & $^{  -0.06}_{   0.06}$ & $^{  -0.02}_{   0.02}$ & $^{   0.02}_{  -0.04}$ &   0.07 \\ 
            &   CaI &    6.34 &    4.61 &  $-$1.73 &    0.19 &  5 &    0.09 & $^{  -0.04}_{   0.04}$ & $^{  -0.01}_{   0.00}$ & $^{   0.02}_{  -0.03}$ &   0.10 \\ 
            &  ScII &    3.15 &    1.02 &  $-$2.13 &  $-$0.21 &  3 &    0.05 & $^{   0.03}_{  -0.04}$ & $^{   0.00}_{  -0.00}$ & $^{   0.06}_{  -0.08}$ &   0.09 \\ 
            &   TiI &    4.95 &    3.12 &  $-$1.83 &    0.09 &  4 &    0.09 & $^{   0.02}_{  -0.02}$ & $^{  -0.00}_{   0.01}$ & $^{  -0.01}_{   0.02}$ &   0.09 \\ 
            &  TiII &    4.95 &    3.23 &  $-$1.72 &    0.20 &  1 &    0.22 & $^{   0.03}_{  -0.04}$ & $^{  -0.00}_{   0.00}$ & $^{   0.01}_{  -0.01}$ &   0.22 \\ 
            &   CrI &    5.64 &    3.61 &  $-$2.03 &  $-$0.11 &  2 &    0.04 & $^{  -0.01}_{   0.00}$ & $^{   0.00}_{  -0.00}$ & $^{   0.05}_{  -0.05}$ &   0.06 \\ 
            &   NiI &    6.22 &    4.21 &  $-$2.02 &  $-$0.10 &  1 &    0.22 & $^{   0.02}_{  -0.01}$ & $^{  -0.01}_{   0.01}$ & $^{  -0.07}_{   0.08}$ &   0.23 \\ 
            &   ZnI &    4.56 &$<$   2.70 &$< -$1.86 &$<$   0.06 &  1 & ... & ... & ... & ... & ... \\ 
            &  SrII &    2.87 &  $-$0.49 &  $-$3.36 &  $-$1.44 &  1 &    1.04 & ...& ...& ...&   1.04 \\ 
            &  BaII &    2.18 &  $-$0.61 &  $-$2.79 &  $-$0.87 &  1 &    0.22 & $^{   0.06}_{  -0.06}$ & $^{   0.01}_{  -0.01}$ & $^{   0.07}_{  -0.09}$ &   0.24 \\ 
\hline
     Boo--911 &   FeI &    7.50 &    5.34 &  $-$2.16 &  $-$2.16 & 26 &    0.02 & $^{   0.17}_{  -0.18}$ & $^{  -0.04}_{   0.05}$ & $^{  -0.12}_{   0.15}$ &   0.23 \\ 
            &  FeII &    7.50 &    5.32 &  $-$2.18 &  $-$2.18 &  2 &    0.06 & $^{  -0.02}_{   0.03}$ & $^{   0.09}_{  -0.08}$ & $^{  -0.12}_{   0.16}$ &   0.18 \\ 
            & C(CH) &    8.43 &    5.50 &  $-$2.93 &  $-$0.77 &  1 &    0.30 & $^{  -0.17}_{   0.18}$ & $^{   0.04}_{  -0.05}$ & $^{   0.12}_{  -0.15}$ &   0.37 \\ 
            &   NaI &    6.24 &    3.80 &  $-$2.44 &  $-$0.28 &  2 &    0.05 & $^{   0.04}_{  -0.05}$ & $^{  -0.03}_{   0.03}$ & $^{  -0.05}_{   0.00}$ &   0.08 \\ 
            &   MgI &    7.60 &    5.79 &  $-$1.81 &    0.35 &  1 &    0.11 & $^{  -0.07}_{   0.08}$ & $^{  -0.02}_{   0.02}$ & $^{   0.00}_{  -0.02}$ &   0.13 \\ 
            &   CaI &    6.34 &    4.17 &  $-$2.17 &  $-$0.01 &  5 &    0.09 & $^{  -0.06}_{   0.06}$ & $^{   0.00}_{  -0.01}$ & $^{   0.04}_{  -0.06}$ &   0.12 \\ 
            &  ScII &    3.15 & ... & ... & ... &  0 & ... & ... & ... & ... & ... \\ 
            &   TiI &    4.95 &    2.60 &  $-$2.35 &  $-$0.19 &  5 &    0.08 & $^{   0.02}_{  -0.03}$ & $^{  -0.00}_{   0.00}$ & $^{   0.05}_{  -0.06}$ &   0.10 \\ 
            &  TiII &    4.95 &    2.79 &  $-$2.16 &    0.02 &  1 &    0.11 & $^{   0.04}_{  -0.04}$ & ...& $^{   0.04}_{  -0.06}$ &   0.13 \\ 
            &   CrI &    5.64 &    2.83 &  $-$2.81 &  $-$0.65 &  1 &    0.11 & $^{   0.03}_{  -0.04}$ & $^{  -0.02}_{   0.02}$ & $^{  -0.07}_{   0.08}$ &   0.14 \\ 
            &   NiI &    6.22 &    3.90 &  $-$2.32 &  $-$0.16 &  3 &    0.17 & $^{  -0.02}_{   0.03}$ & $^{   0.01}_{  -0.01}$ & $^{   0.03}_{  -0.04}$ &   0.17 \\ 
            &   ZnI &    4.56 &$<$   2.20 &$< -$2.36 &$< -$0.20 &  1 & ... & ... & ... & ... & ... \\ 
            &  SrII &    2.87 & ... & ... & ... &  0 & ... & ... & ... & ... & ... \\ 
            &  BaII &    2.18 &  $-$0.64 &  $-$2.82 &  $-$0.64 &  3 &    0.12 & $^{   0.08}_{  -0.08}$ & $^{   0.01}_{  -0.01}$ & $^{  -0.02}_{   0.03}$ &   0.15 \\ 
\hline
    HD216143 &   FeI &    7.50 &    5.35 &  $-$2.15 &  $-$2.15 & 64 &    0.01 & $^{   0.15}_{  -0.16}$ & $^{  -0.03}_{   0.03}$ & $^{  -0.07}_{   0.09}$ &   0.18 \\ 
            &  FeII &    7.50 &    5.35 &  $-$2.15 &  $-$2.15 & 13 &    0.01 & $^{  -0.02}_{   0.03}$ & $^{   0.11}_{  -0.10}$ & $^{  -0.04}_{   0.05}$ &   0.12 \\ 
            & C(CH) &    8.43 & ... & ... & ... &  0 & ... & ... & ... & ... & ... \\ 
            &   NaI &    6.24 & ... & ... & ... &  0 & ... & ... & ... & ... & ... \\ 
            &   MgI &    7.60 &    5.80 &  $-$1.80 &    0.35 &  2 &    0.07 & $^{  -0.05}_{   0.05}$ & $^{  -0.04}_{   0.04}$ & $^{  -0.03}_{   0.02}$ &   0.10 \\ 
            &   CaI &    6.34 &    4.40 &  $-$1.94 &    0.21 & 16 &    0.03 & $^{  -0.05}_{   0.05}$ & $^{  -0.00}_{  -0.00}$ & $^{   0.03}_{  -0.04}$ &   0.07 \\ 
            &  ScII &    3.15 &    0.99 &  $-$2.16 &  $-$0.01 &  8 &    0.03 & $^{   0.04}_{  -0.04}$ & $^{  -0.00}_{   0.00}$ & $^{   0.01}_{  -0.01}$ &   0.05 \\ 
            &   TiI &    4.95 &    2.92 &  $-$2.04 &    0.11 & 18 &    0.03 & $^{   0.02}_{  -0.03}$ & $^{  -0.01}_{   0.00}$ & $^{   0.03}_{  -0.04}$ &   0.05 \\ 
            &  TiII &    4.95 &    3.00 &  $-$1.95 &    0.20 &  6 &    0.04 & $^{   0.03}_{  -0.04}$ & $^{  -0.01}_{   0.01}$ & $^{  -0.04}_{   0.05}$ &   0.07 \\ 
            &   CrI &    5.64 &    3.23 &  $-$2.41 &  $-$0.26 &  4 &    0.03 & $^{   0.01}_{  -0.01}$ & $^{  -0.01}_{   0.01}$ & $^{   0.02}_{  -0.02}$ &   0.04 \\ 
            &   NiI &    6.22 &    4.02 &  $-$2.20 &  $-$0.05 & 12 &    0.02 & $^{  -0.04}_{   0.05}$ & $^{   0.02}_{  -0.02}$ & $^{   0.05}_{  -0.07}$ &   0.08 \\ 
            &   ZnI &    4.56 &    2.55 &  $-$2.01 &    0.14 &  2 &    0.01 & $^{  -0.13}_{   0.16}$ & $^{   0.09}_{  -0.09}$ & $^{   0.04}_{  -0.05}$ &   0.17 \\ 
            &  SrII &    2.87 & ... & ... & ... &  0 & ... & ... & ... & ... & ... \\ 
            &  BaII &    2.18 &  $-$0.09 &  $-$2.27 &  $-$0.12 &  3 &    0.06 & $^{   0.07}_{  -0.09}$ & $^{  -0.01}_{   0.01}$ & $^{  -0.13}_{   0.16}$ &   0.18 \\ 
\hline
     HD85773 &   FeI &    7.50 &    5.11 &  $-$2.39 &  $-$2.39 & 58 &    0.01 & $^{   0.14}_{  -0.16}$ & $^{  -0.04}_{   0.04}$ & $^{  -0.06}_{   0.07}$ &   0.17 \\ 
            &  FeII &    7.50 &    5.17 &  $-$2.33 &  $-$2.33 & 12 &    0.01 & $^{  -0.02}_{   0.03}$ & $^{   0.10}_{  -0.10}$ & $^{  -0.03}_{   0.04}$ &   0.11 \\ 
            & C(CH) &    8.43 & ... & ... & ... &  0 & ... & ... & ... & ... & ... \\ 
            &   NaI &    6.24 &    3.99 &  $-$2.25 &    0.14 &  1 &    0.11 & $^{  -0.08}_{   0.09}$ & $^{   0.02}_{  -0.02}$ & $^{   0.05}_{  -0.07}$ &   0.15 \\ 
            &   MgI &    7.60 &    5.45 &  $-$2.15 &    0.24 &  3 &    0.08 & $^{  -0.07}_{   0.07}$ & $^{  -0.02}_{   0.01}$ & $^{  -0.00}_{  -0.01}$ &   0.10 \\ 
            &   CaI &    6.34 &    4.11 &  $-$2.23 &    0.16 & 16 &    0.02 & $^{  -0.05}_{   0.05}$ & $^{   0.00}_{  -0.01}$ & $^{   0.02}_{  -0.03}$ &   0.07 \\ 
            &  ScII &    3.15 &    0.74 &  $-$2.41 &  $-$0.08 &  6 &    0.03 & $^{   0.04}_{  -0.05}$ & $^{  -0.00}_{   0.00}$ & $^{   0.01}_{  -0.01}$ &   0.06 \\ 
            &   TiI &    4.95 &    2.73 &  $-$2.22 &    0.17 & 19 &    0.02 & $^{   0.03}_{  -0.05}$ & $^{  -0.01}_{   0.00}$ & $^{   0.01}_{  -0.01}$ &   0.05 \\ 
            &  TiII &    4.95 &    2.89 &  $-$2.06 &    0.28 &  6 &    0.04 & $^{   0.04}_{  -0.04}$ & $^{  -0.02}_{   0.02}$ & $^{  -0.05}_{   0.07}$ &   0.08 \\ 
            &   CrI &    5.64 &    3.03 &  $-$2.61 &  $-$0.22 &  4 &    0.02 & $^{   0.01}_{  -0.02}$ & $^{  -0.01}_{   0.01}$ & $^{   0.01}_{  -0.01}$ &   0.03 \\ 
            &   NiI &    6.22 &    3.85 &  $-$2.37 &    0.02 & 11 &    0.02 & $^{  -0.04}_{   0.05}$ & $^{   0.02}_{  -0.02}$ & $^{   0.04}_{  -0.05}$ &   0.07 \\ 
            &   ZnI &    4.56 &    2.67 &  $-$1.89 &    0.50 &  2 &    0.00 & $^{  -0.13}_{   0.17}$ & $^{   0.09}_{  -0.09}$ & $^{   0.01}_{  -0.01}$ &   0.17 \\ 
            &  SrII &    2.87 & ... & ... & ... &  0 & ... & ... & ... & ... & ... \\ 
            &  BaII &    2.18 &  $-$0.68 &  $-$2.86 &  $-$0.53 &  3 &    0.07 & $^{   0.09}_{  -0.09}$ & $^{  -0.01}_{   0.01}$ & $^{  -0.08}_{   0.11}$ &   0.14 \\ 
\hline
\hline
\end{longtable}

\end{document}